\documentclass[twocolumn,twoside,slac]{revtex4}
\usepackage{graphicx}
\usepackage{fancyhdr}
\usepackage{epsfig}
\begin{document}

\title{Physics Analysis Expert PAX: First Applications}

\author{Martin Erdmann$^1$,
Dominic Hirschb\"uhl, 
Christopher Jung, 
Steffen Kappler$^2$,
Yves Kemp, 
Matthias Kirsch, 
Deborah Miksat, 
Christian Piasecki, 
G\"unter Quast,
Klaus Rabbertz,
Patrick Schemitz, 
Alexander Schmidt, 
Thorsten Walter,
Christian Weiser}
\affiliation{
Institut f\"ur Experimentelle Kernphysik,
Universit\"at Karlsruhe, Wolfgang-Gaede-Str.\ 1, \\
D-76131 Karlsruhe, Germany \\
$^1${\rm contact} Martin.Erdmann@cern.ch \\
$^2${\rm also with } CERN, EP Division, CH-1211 Geneva 23, Switzerland
}

\begin{abstract}
PAX (\underline{P}hysics \underline{A}nalysis E\underline{x}pert) 
is a novel, C++ based toolkit designed to assist teams in particle 
physics data analysis issues. 
The core of PAX are event interpretation containers, holding relevant 
information about and possible interpretations of a physics event. 
Providing this new level of abstraction beyond the results of the 
detector reconstruction programs, PAX facilitates the buildup and 
use of modern analysis factories. Class structure and user command
syntax of PAX are set up to support expert teams as well as newcomers in 
preparing for the challenges expected to arise in the data analysis 
at future hadron colliders. 
\end{abstract}

\maketitle

\thispagestyle{fancy}

\section{Motivation}

\noindent
Working directly on the output of detector reconstruction programs
when performing data analyses is an established habit amongst particle 
physicists. Nevertheless, at the experiments of HERA and LEP it 
turned out to be an advantage to have uniform access to 
calorimeter energy depositions, tracks, electrons, muons etc.\ which
requires a new level of abstraction on top of the reconstruction 
layer. Examples of physics analysis packages providing this level are 
H1PHAN\footnote{H1 Collaboration, internal software manual for H1PHAN}
and ALPHA\footnote{ALEPH Collaboration, ``ALPHA'' internal note 99-087 SOFTWR 99-001} 
of the H1 and ALEPH 
experiments. Noticed effects were, amongst others, that \\
a) users could relatively quickly answer physics questions,\\
b) the physics analysis code was protected against changes in the
   detector reconstruction layer,\\
c) and finally the management liked the fact, that the relevant
   reconstruction output had been used in the analysis.

\begin{figure*}
\setlength{\unitlength}{1cm}
\begin{picture}(16.5,5.0)
\put(0,1){\epsfig{file=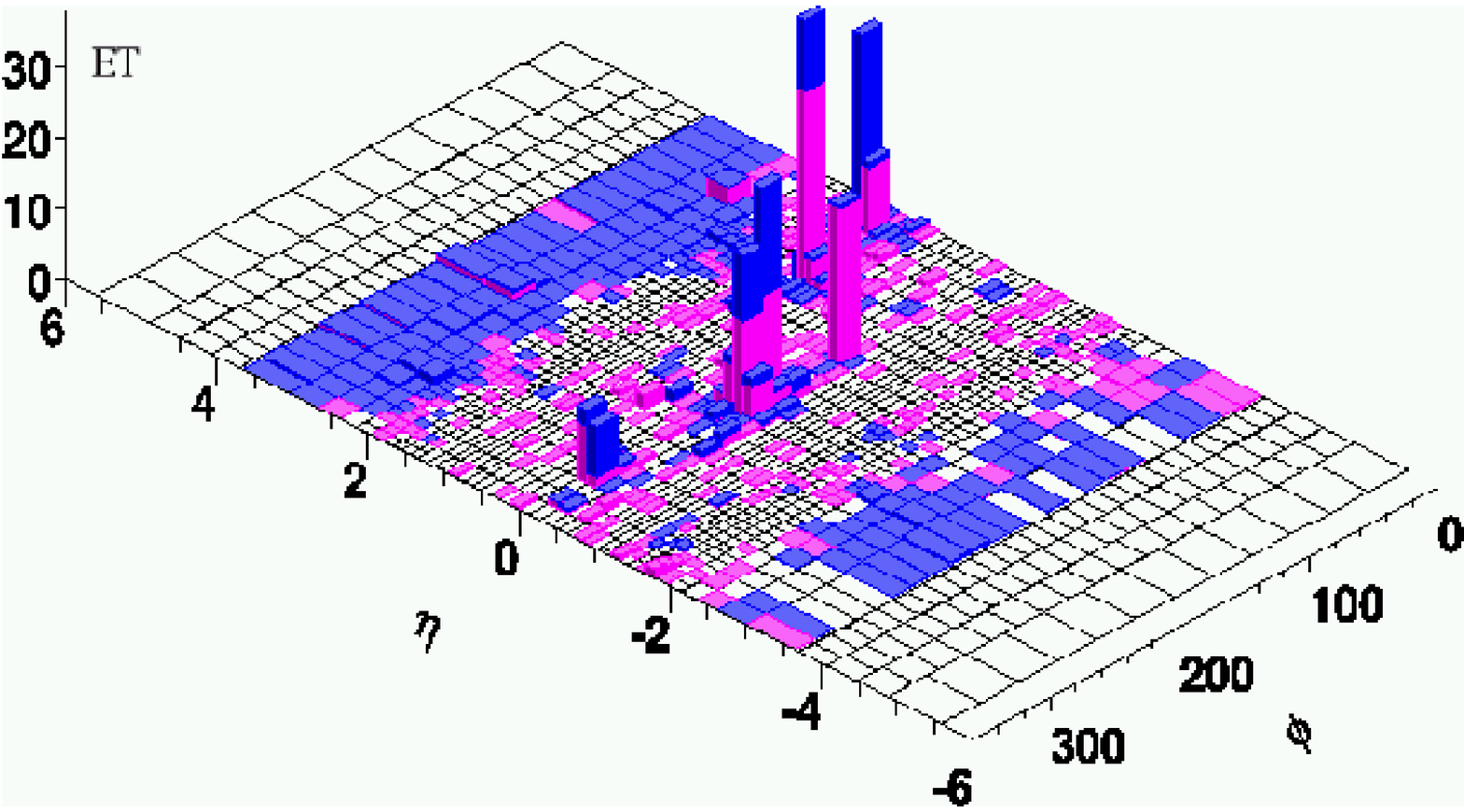,width=6cm}}
\put(8,3.){\epsfig{file=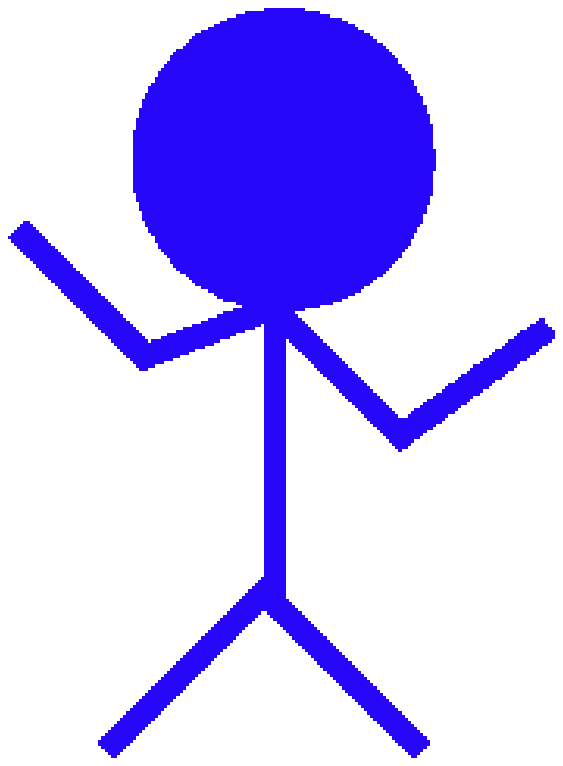,width=1.5cm}}
\put(6.5,0.){\epsfig{file=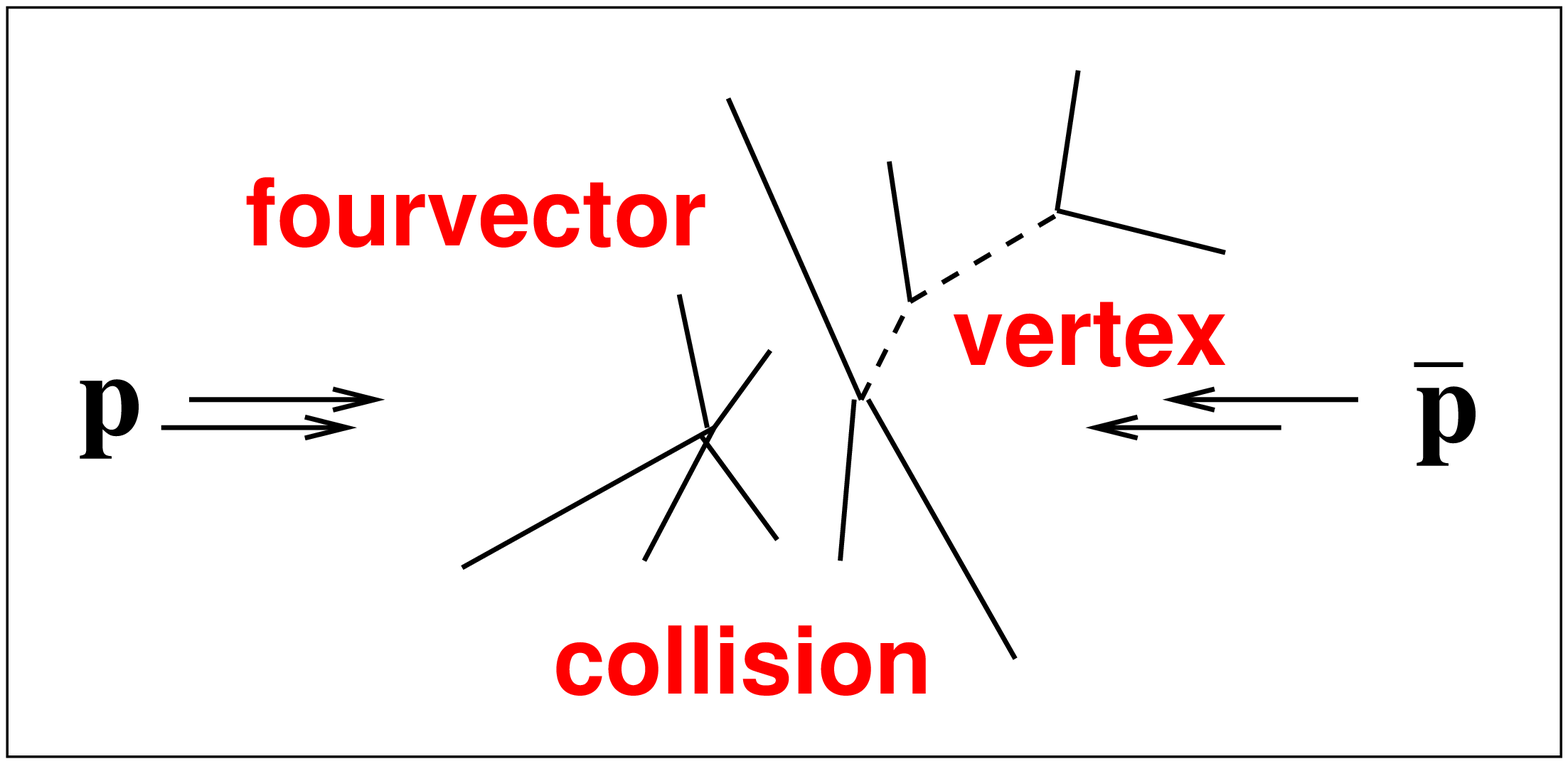,width=5cm}}
\put(12.5,0.5){\epsfig{file=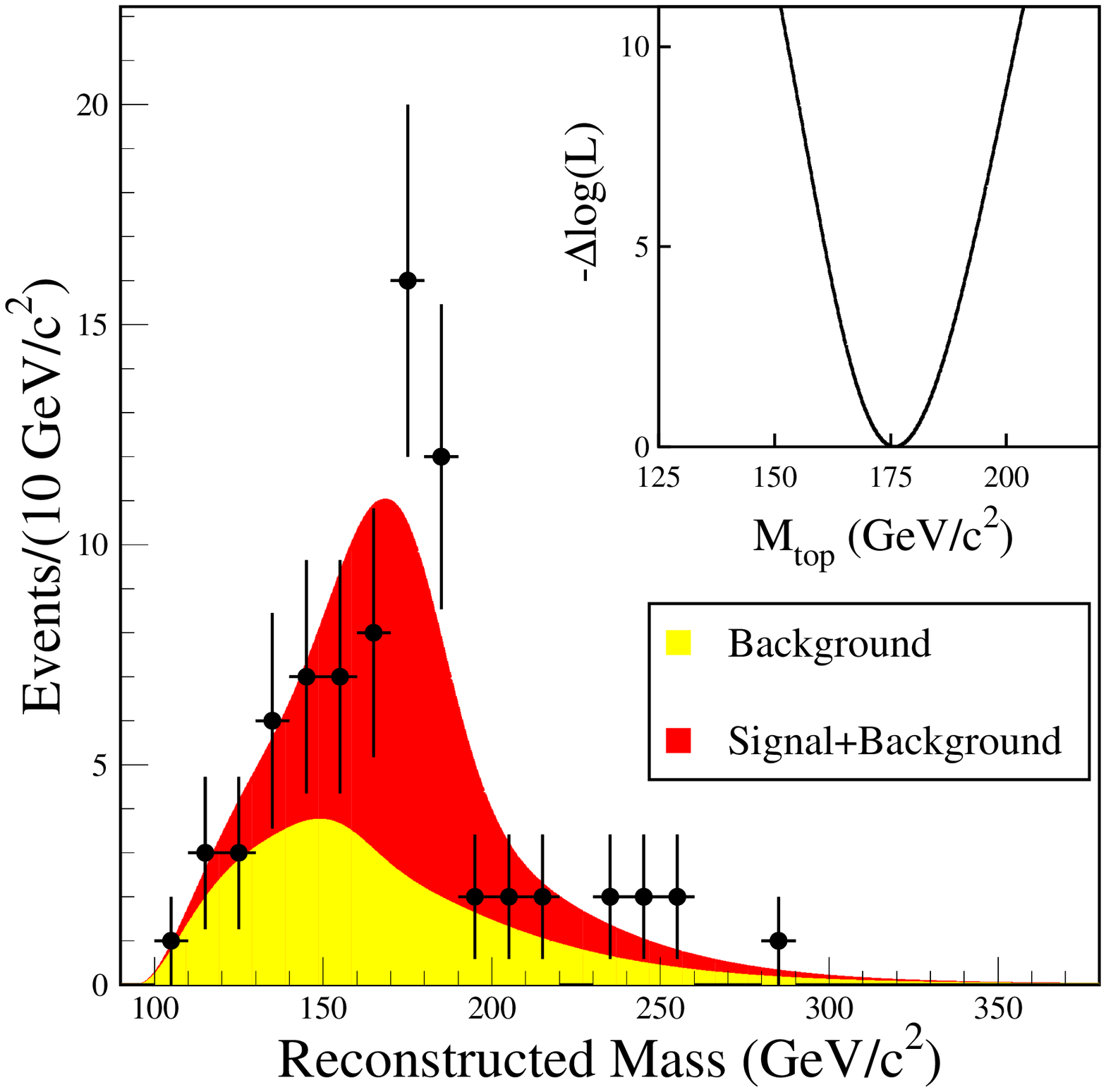,width=4cm}}
\end{picture}
\caption{ 
Application area of the PAX toolkit:
in between detector reconstruction output and 
physics interpretation of an event \protect{\cite{cdftop}}.
\label{fig:pax}}
\end{figure*}

While previous programs were designed to provide a rather complete view 
of the event originating from a single $e^+e^-$ or $ep$ scattering, a next 
generation package is challenged by hadron collisions with $O(20)$
simultaneous events. This implies a large number of possible 
interpretations of the triggered events and sometimes requires the 
analysis of dedicated regions of interest. 
The ``\underline{P}hysics \underline{A}nalysis E\underline{x}pert'' 
toolkit (PAX) is a data analysis utility designed to assist physicists in 
these tasks in the phase between detector reconstruction and physics 
interpretation of an event (Fig.\ref{fig:pax}).
The alpha-version of PAX was presented at the HCP2002 conference \cite{alfa}.
In this contribution we introduce the beta-version.

\section{Guidelines for the PAX Design}

\noindent
The design of the next generation physics analysis utility PAX has been 
developed according to the guidelines listed below:
\begin{enumerate}
\item
The package is a utility tool box in a sense that the user has full 
control of every step in the program execution. 
\item
The programming interface should be as simple and intuitive as possible. 
This minimizes the need to access the manual and thereby increases the 
acceptance in the community. Furthermore, simplicity enables also physicists 
with limited time budget or limited knowledge of object oriented programming to 
carry out complex physics analyses.
\item
The package supports modular physics analysis structures and thus facilitates 
team work. The complexity of todays and future analyses makes team work
of many physicists mandatory.
\item
Existing physics analysis utilities can be connected.
Examples are tools for fourvector gymnastics which are available
in general form (e.g.\ in the CLHEP 
library\footnote{CLHEP, A Class Library for High Energy Physics,
{\em http://proj-clhep.web.cern.ch/proj-clhep/ }}), 
other examples are 
histograms, fitting routines etc.
\item
The physics analysis package can be used consistently among different 
high energy physics experiments.
\item
Many use cases are to be taken care of.
The following list is certainly not complete:
\begin{enumerate}
\item
Access to the original data of the experiment is possible at each 
stage of the physics analysis.
This enables detailed control of all
analysis steps, and access to experiment-dependent 
information and related methods.
\item
When studying events of a Monte Carlo generator,
relations between generated and reconstructed observables 
are accessible at any stage of the analysis.
This allows the quality of the detector reconstruction
to be studied in detail.
\item
Without significant code changes, a complete analysis chain can be 
tested with different input objects such as reconstructed tracks, 
generated particles, fast simulated particles etc. 
\item
Relations between reconstructed physics objects (tracks, muons, etc.)
and vertices are available, as well as hooks for separating
multiple interactions.
\item
The decay chains with secondary, tertiary etc.\ vertices can be
handled in events with multiple interactions.
\item
Information of different objects can be combined, e.g., 
tracks and calorimeter information.
\item
A common challenge in data analysis are reconstruction ambiguities 
which need to be handled.
Administration of these ambiguities is supported.
\item
The user finds assistance in developing analysis factories with
multiple physics data analyses carried out simultaneously.
\end{enumerate}
\end{enumerate}
To cope with these challenges, the advantage of using an object 
oriented language is obvious. For the convenience of connecting to other
packages, C++ was the language of choice for the realisation of PAX. 

\section{PAX Class Structure Implementation}

\subsection{Event Interpretation}

\begin{figure}[t]
\setlength{\unitlength}{1cm}
\begin{picture}(10.0,6)
\put(0,0){\epsfig{file=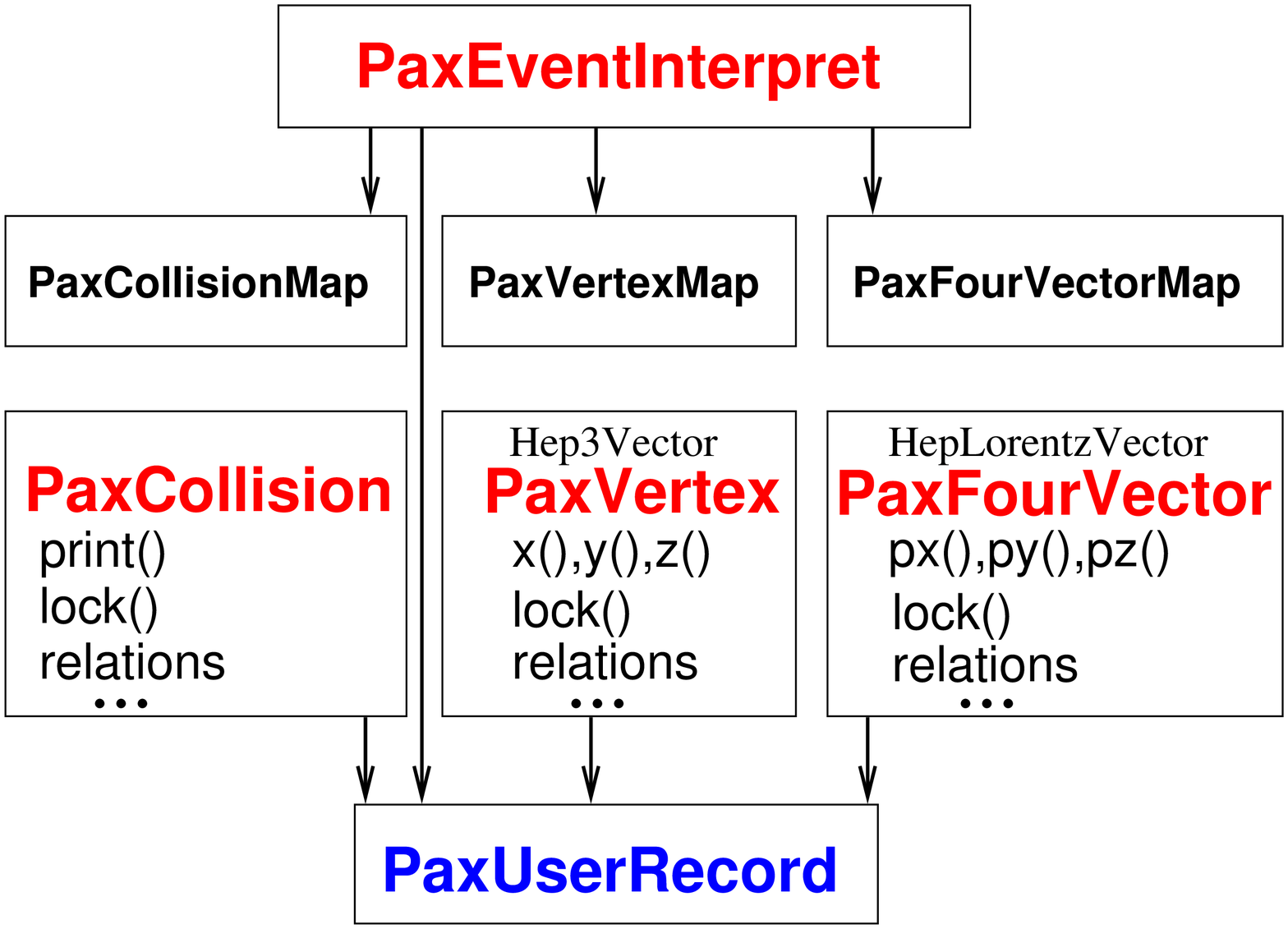,width=8cm}}
\end{picture}
\caption{ The basic unit in PAX: the event interpretation
together with the classes for collisions, vertices, fourvectors, and 
user records.
\label{fig:eventinterpret}}
\end{figure}

\noindent
The basic unit in PAX is a view of the event which we call ``event interpretation''.
The event interpretation is a container used to store the relevant 
information about the event in terms of collisions, vertices, fourvectors, 
their relations, and additional values needed in the user's analysis.
The corresponding class is denoted {\em PaxEventInterpret}  
(Fig.\ref{fig:eventinterpret}).
The user books, fills, draws, copies, advances the event interpretation,
and has the ultimate responsibility for deleting it.
When the user finally deletes an instance of {\em PaxEventInterpret},
instances of objects which have been registered with this event 
interpretation  -- collisions, vertices, fourvectors, etc.\ -- are also 
removed from memory. 

A copy of an event interpretation is a physical copy in memory.
It is generated preserving all values of the original event interpretation,
and with all relations between collisions, vertices, and fourvectors 
corrected to remain within the copied event interpretation.
In addition, the histories of the individual collisions, vertices,
and fourvectors are recorded.
The copy functionality simplifies producing several similar event interpretations.
This is advantageous typically in the case of many possible 
views of the event that differ in a few aspects only. 
Although the event interpretations do not know of each others existence,
recording the analysis history of collisions, vertices and fourvectors 
requires intermediate event interpretations to exist. 
Therefore, we recommend to delete all event 
interpretations together after an event has been analysed.

Besides the features mentioned, the {\em PaxEventInterpret} also defines an interface 
to algorithms such as jet algorithms, missing transverse energy calculations etc. 
This eases the exchange of algorithms within, or between analysis teams. 

\subsection{Physics Quantities}

\noindent
PAX supports three physics quantities: collisions, vertices, and fourvectors.
Three classes have been defined correspondingly.
The class {\em PaxCollision} provides the hooks to handle multi-collision events
(Fig.\ref{fig:eventinterpret}).
Vertices and fourvectors are defined through the classes {\em PaxVertex}, and 
{\em PaxFourVector}.
Since the user may need to impose vector operations on them,
both PAX classes inherit from the CLHEP classes {\em Hep3Vector},
and {\em HepLorentzVector}, respectively.
Their functionalities are available to the user.
The {\em PaxVertex} and {\em PaxFourVector} classes 
contain additional functionality
which mainly result from features proven to be useful in the
previously mentioned H1PHAN package.

For all physics quantities the user can store additional values (data type 
{\em double}) needed by the analysis via the class {\em PaxUserRecord}.
These values are registered together with a key (data type {\em string} functioning
as a name), which must be given by the user.

\subsection{Relation Management for the Physics Quantities}

\begin{figure}[hhh]
\setlength{\unitlength}{1cm}
\begin{picture}(10.0,6)
\put(0,0)
{\epsfig{file=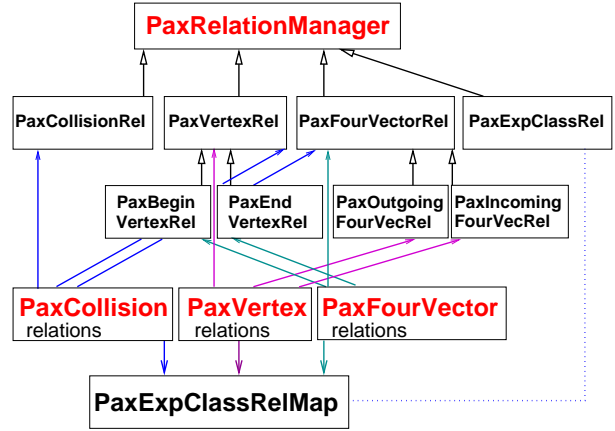,width=8cm}}
\end{picture}
\caption{ The relations in PAX enable storage of decay trees,
records of analysis history,
access to experiment specific classes, 
and exclusion of parts of the event from the analysis.
\label{fig:relation}}
\end{figure}

\noindent
Relations between collisions, vertices, and fourvectors are handled by
a separate class called {\em PaxRelationManager} (Fig.\ref{fig:relation}).
Here we followed the design pattern {\em Mediator} \cite{designpatterns}.
Examples for relations to be handled between physics quantities are 
fourvectors which originate from the primary vertex, 
an incoming fourvector to a secondary vertex, or
connections between multiple collisions and their vertices etc.

When physics quantities are copied, the copied instance carries a
pointer to the previous instance.
An example would be a fourvector which is copied together with an event 
interpretation.
In this way, the full history of the fourvector is kept throughout
the analysis.

\begin{figure}[hhh]
\setlength{\unitlength}{1cm}
\begin{picture}(10.0,3.5)
\put(0.5,0.)
{\epsfig{file=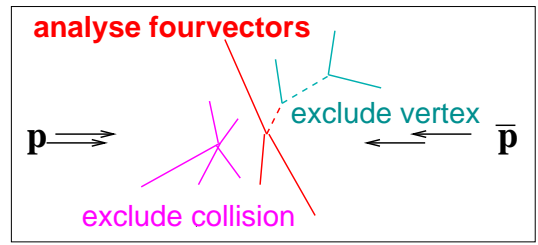,width=7cm}}
\end{picture}
\caption{ Excluding a collision (left) or a vertex (right) from an analysis
using the lock mechanism excludes all vertices and fourvectors originating
from the excluded object.
\label{fig:lock}}
\end{figure}

The relation manager also allows parts of the event to be excluded
from the analysis.
An example would be a lepton which needs to be preserved while applying 
a jet algorithm.
This locking mechanism is build in the form of a tree structure
which enables sophisticated exclusion of unwanted event parts.
For example, locking a collision excludes the vertices and fourvectors
connected to this collision (Fig.\ref{fig:lock}).
In the case of locking a secondary vertex, PAX will lock the decay 
tree starting at this vertex.

Note that when locking a fourvector $f$, all fourvectors related to $f$ 
through its history record are locked as well.
The same functionality applies to collisions and vertices.
Unlocking the fourvector $f$ removes the lock flag of all physics
quantities related to $f$ through the decay tree, or the history record
at the time the unlock command is executed.

For some applications, the user may want to inquire additional information 
on a physics quantity which is only contained in an experiment specific class.
An example is a {\em PaxFourVector} instance originating from a track 
of which the user wants to check the $\chi^2$ probability of the track fit.
The relation manager allows to register instances of experiment specific 
classes which led to a {\em PaxCollision}, {\em PaxVertex}, or {\em PaxFourVector}
instance.
To enable such relations, a template class {\em PaxExperiment$<>$} 
has been defined which allows registration of arbitrary class instances.
Applying the C++ {\em dynamic$\_$cast} operator, the user can recover the 
original instance, and access all member functions of the experiment 
specific class.

\subsection{Container and Iterator}

\noindent
PAX uses the template class {\em map$<>$} from the Standard Template Library (STL) 
\cite{STL}
to manage pairs of keys and items in a container.
The user record in Fig.\ref{fig:eventinterpret} is an example of such a container for 
pairs of data type {\em string} and {\em double}. 
For accessing a certain item, optimized STL 
algorithms search the {\em map} for the corresponding key and provide access to the 
item.

All {\em PaxCollision}, {\em PaxVertex}, and {\em PaxFourVector} instances carry 
a unique identifier of type {\em PaxId} which is used as the key in the 
{\em PaxCollisionMap}, 
{\em PaxVertexMap}, and {\em PaxFourVectorMap} of an event interpretation 
(Fig.\ref{fig:eventinterpret}). Pointers to the collision, vertex, and fourvector 
instances are the corresponding items. 
In this way, fast and uniform access to the 
individual physics quantities is guaranteed.

For users not familiar with STL iterators, we provide the {\em PaxIterator} class 
which gives a simple and unified command syntax for accessing all containers in PAX.

\subsection{Documentation}

\noindent
The PAX user guide is available on the web \cite{webnavigator}.
In addition to a paper version of the manual, we provide a fast 
navigator web page which can be used as a reference guide.

\section{Application within Physics Analysis of the CDF Experiment}

\begin{figure}
\setlength{\unitlength}{1cm}
\begin{picture}(10.0,7.0)
\put(0,3.5){\epsfig{file=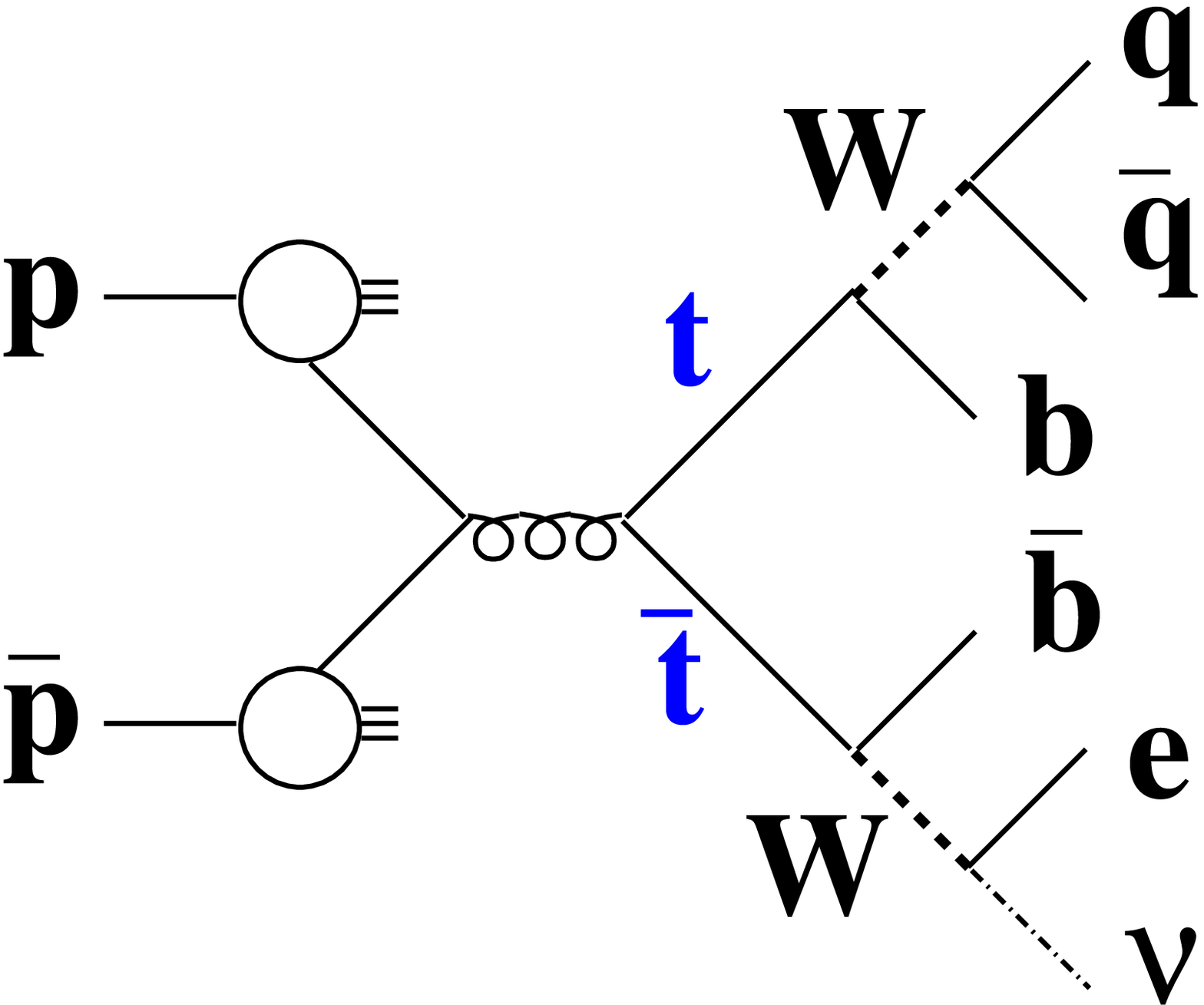,width=3.8cm}}
\put(4.5,3.5){\epsfig{file=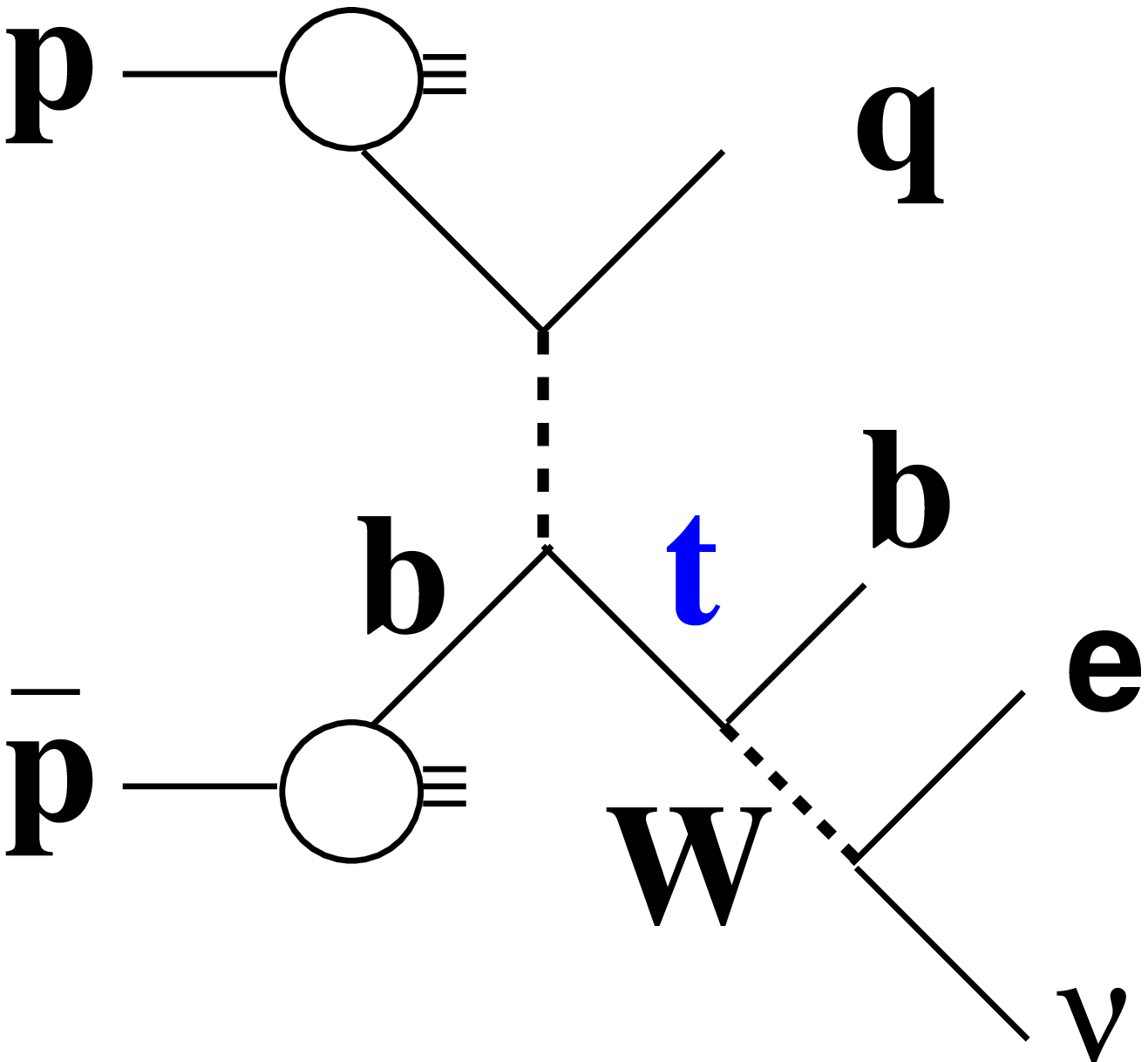,width=3.3cm}}
\put(0,0){\epsfig{file=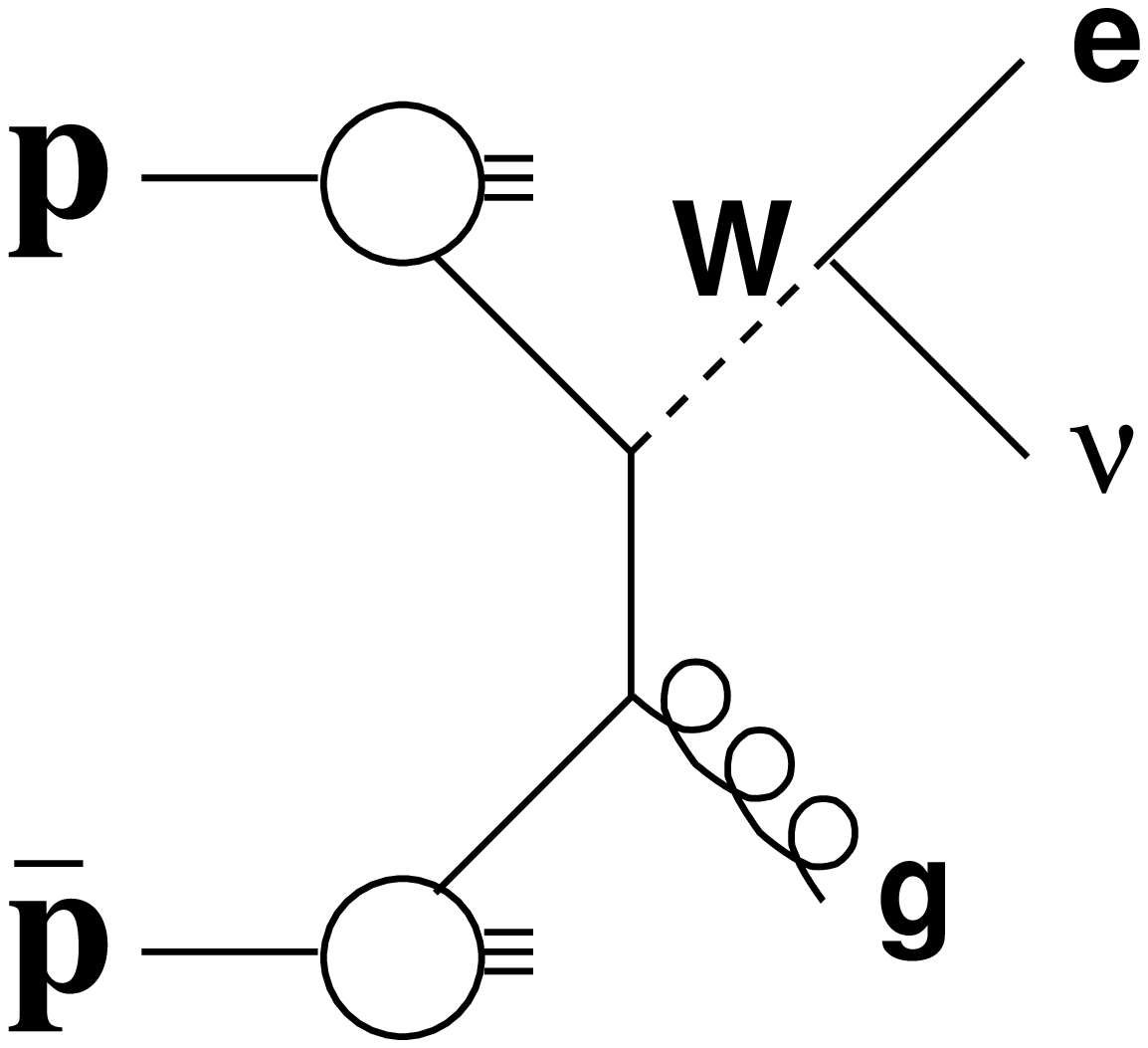,width=3.3cm}}
\put(4.5,0){\epsfig{file=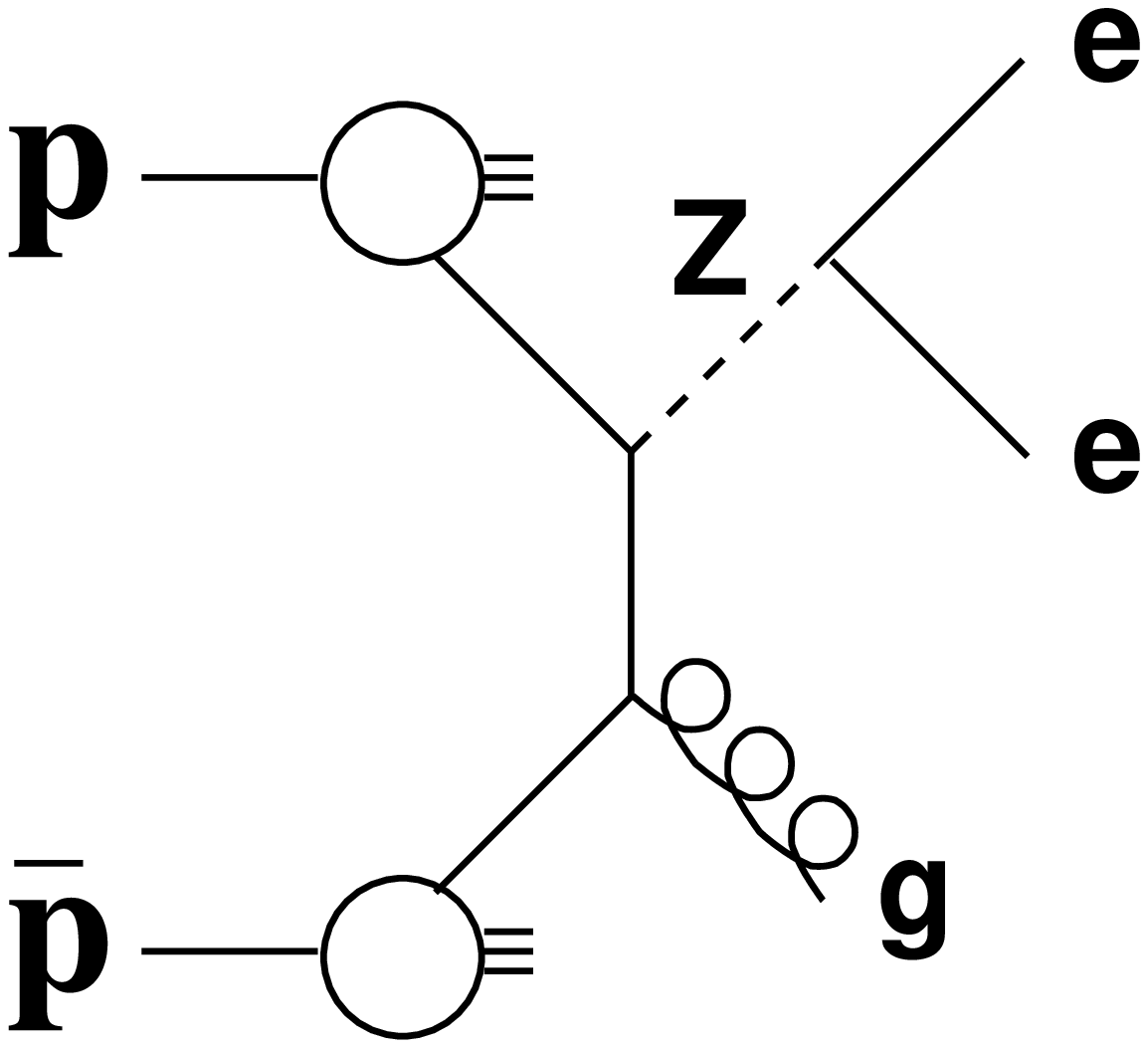,width=3.3cm}}
\end{picture}
\caption{Example Feynman diagrams relevant to top quark analysis 
in proton-antiproton collisions.
\label{fig:feynman_top}}
\end{figure}

\noindent
The PAX package is explored by the Karlsruhe CDF group in top quark analyses.
Example Feynman diagrams of signal and background processes relevant to
top analysis in the so-called electron plus jet channel are shown in 
Fig.\ref{fig:feynman_top}.

\subsection{Combining Results of the Detector Reconstruction Program}

\begin{figure}
\setlength{\unitlength}{1cm}
\begin{picture}(10,16)
\put(1,12.5){\epsfig{file=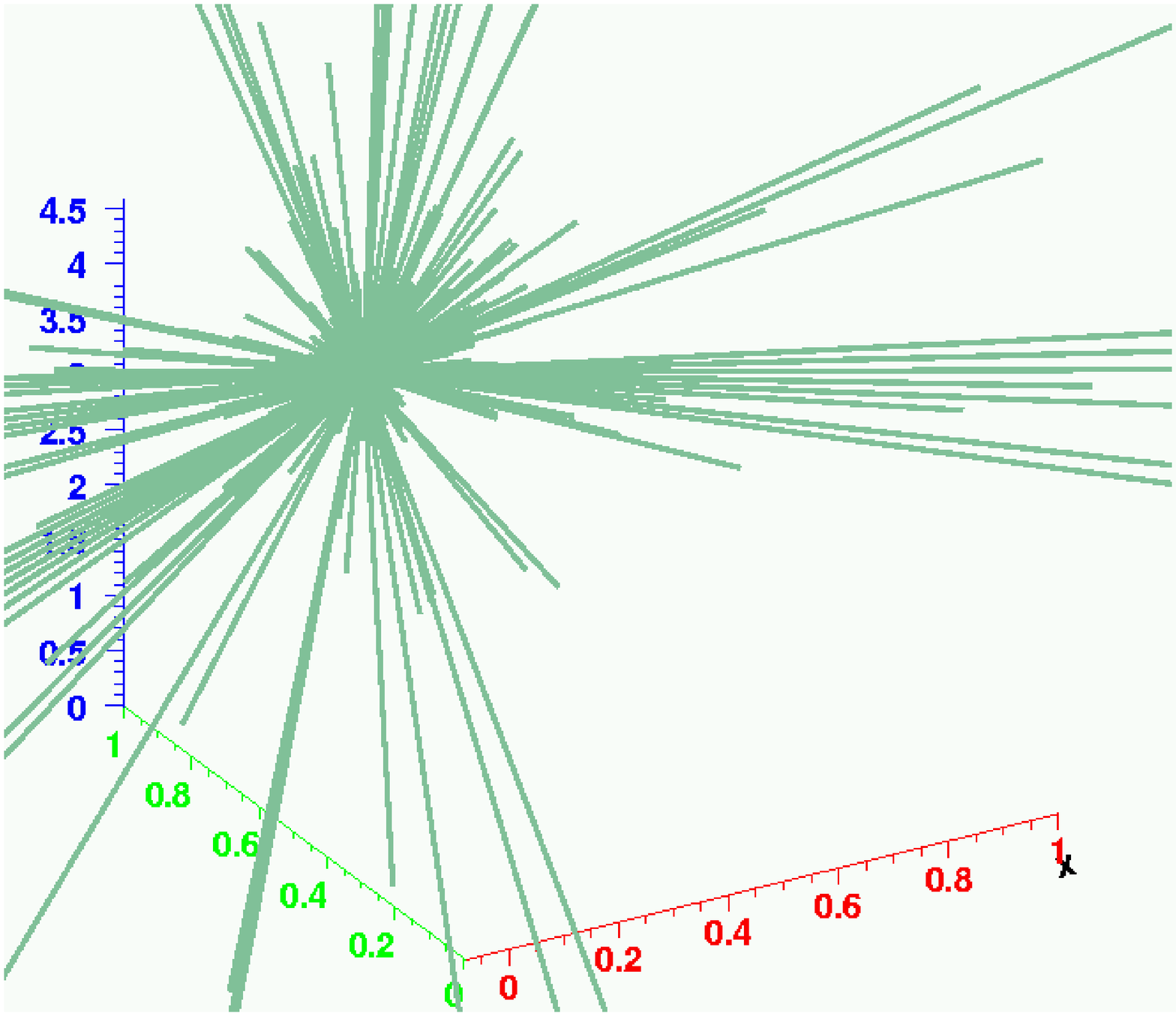,width=3cm}}
\put(4.5,12.5){\epsfig{file=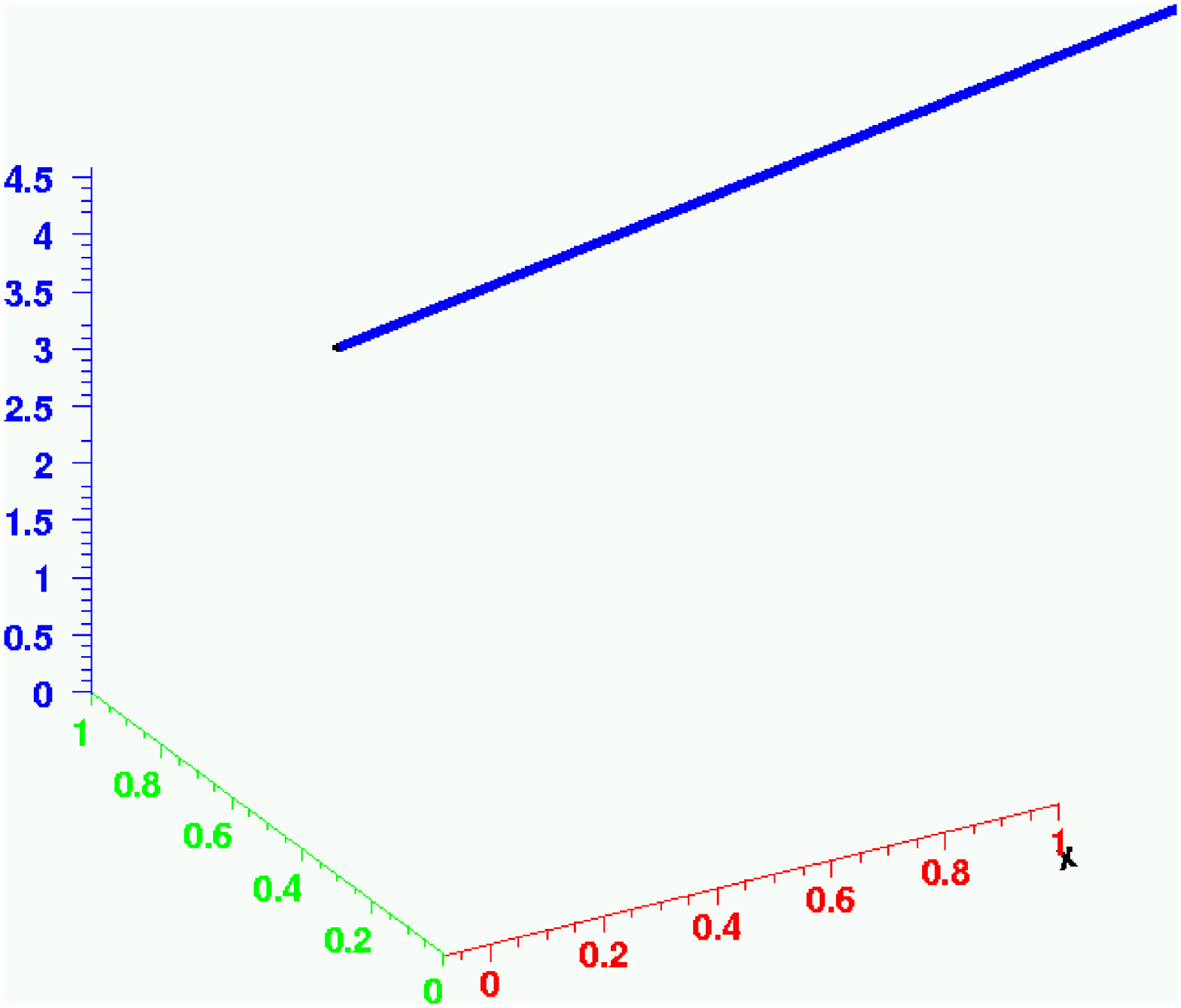,width=3cm}}
\put(0,9){\epsfig{file=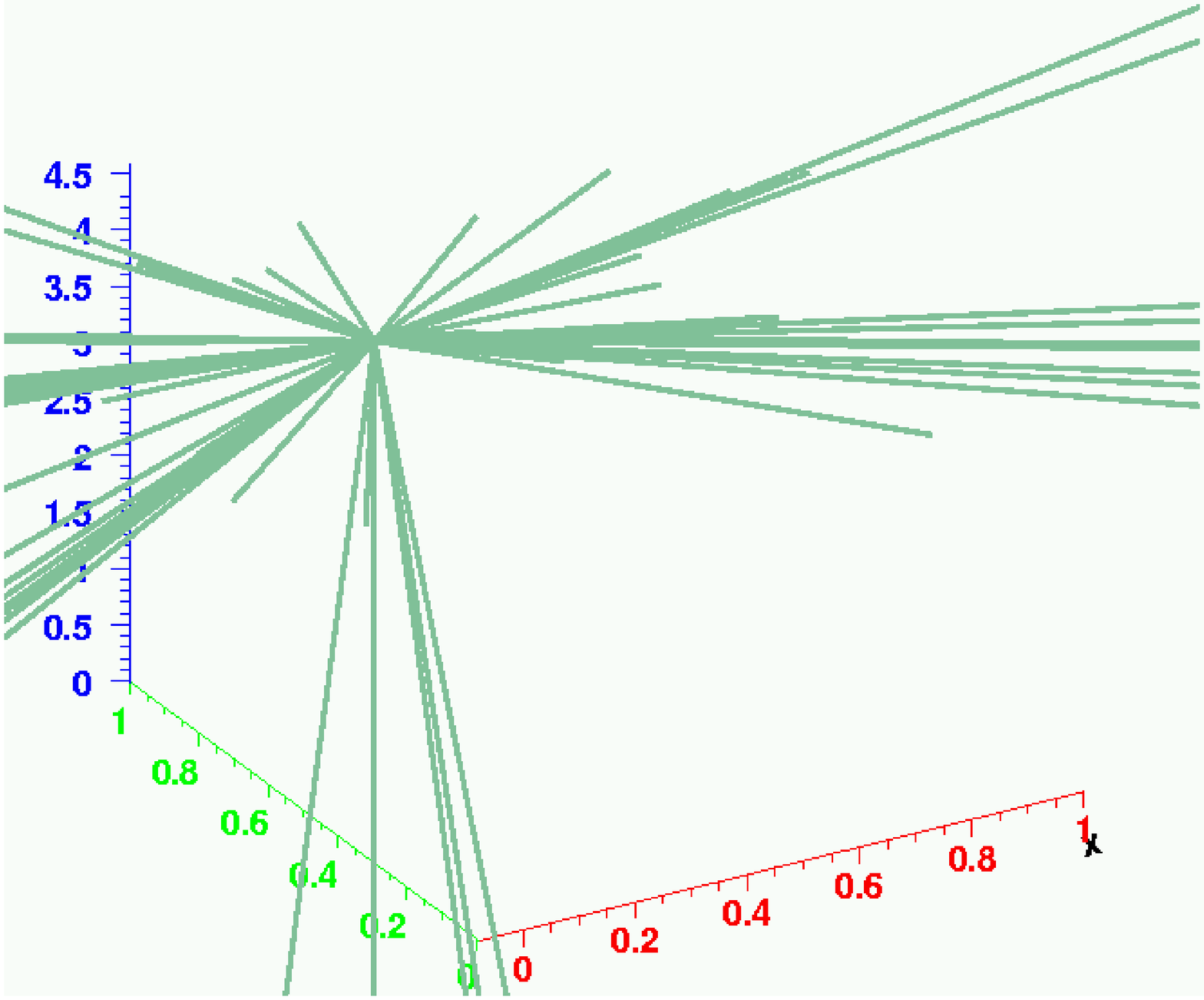,width=3cm}}
\put(3.5,8){\epsfig{file=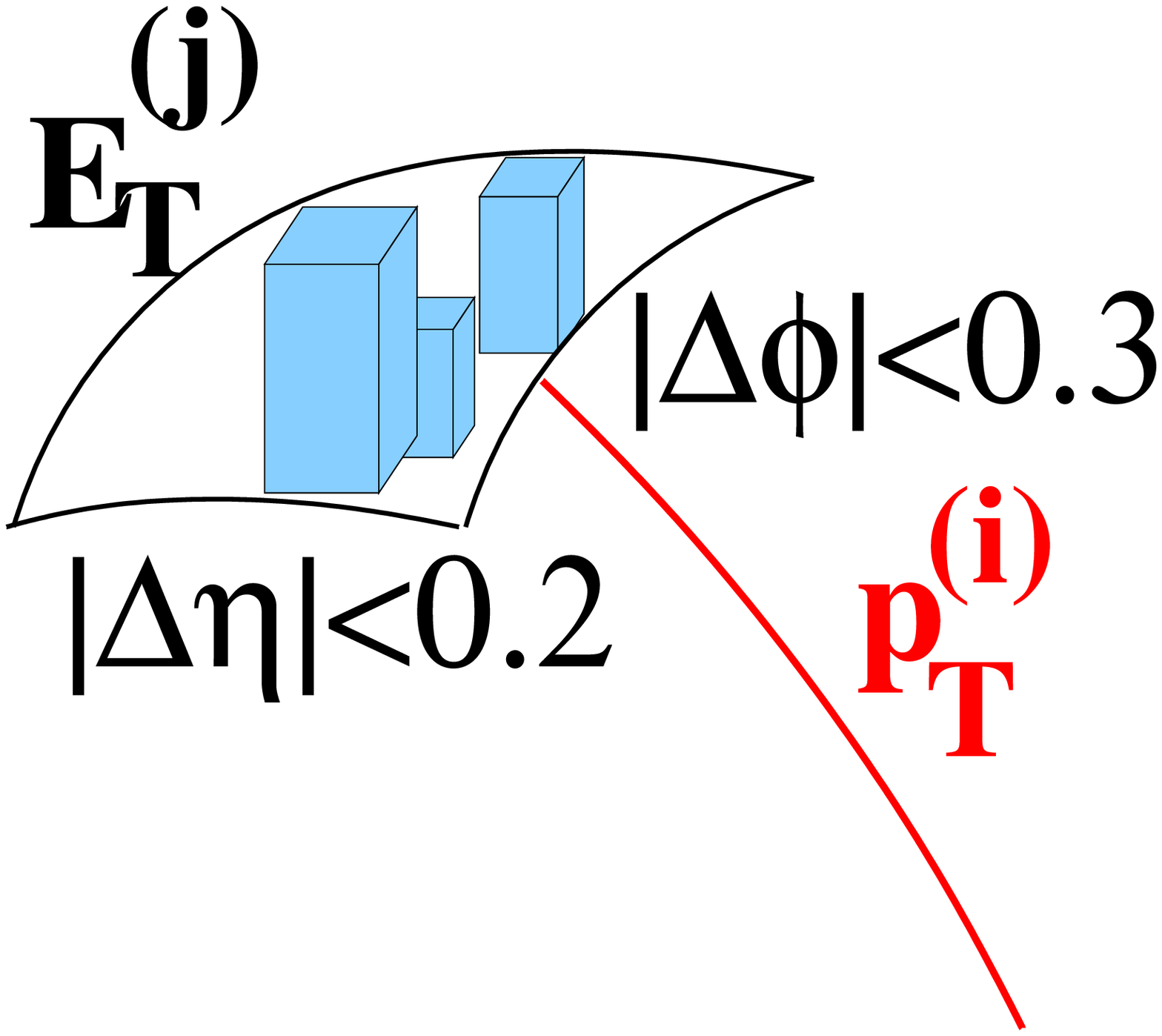,width=4cm}}
\put(0.5,0){\epsfig{file=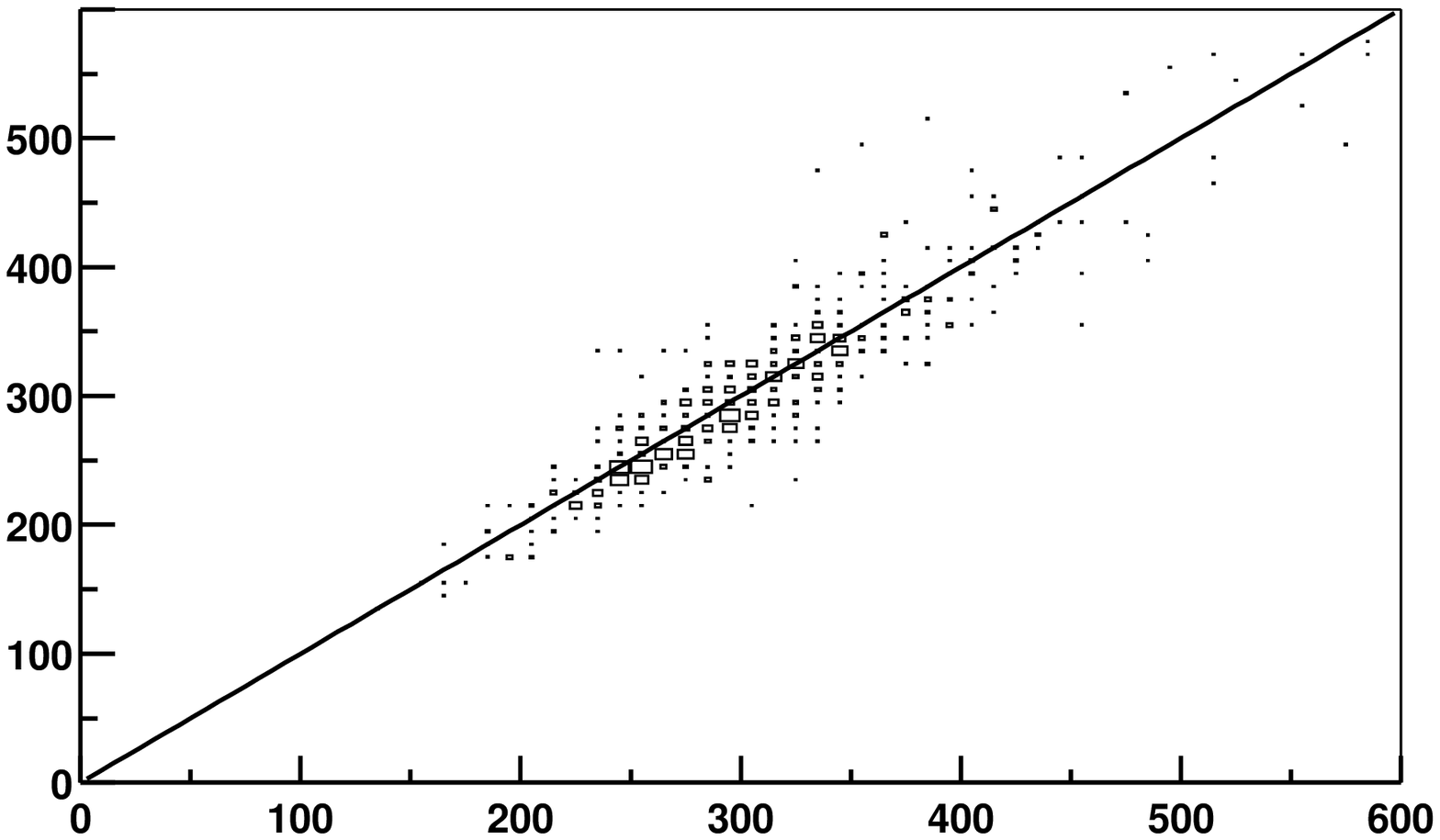,width=7.5cm}}
\put(4,5){\epsfig{file=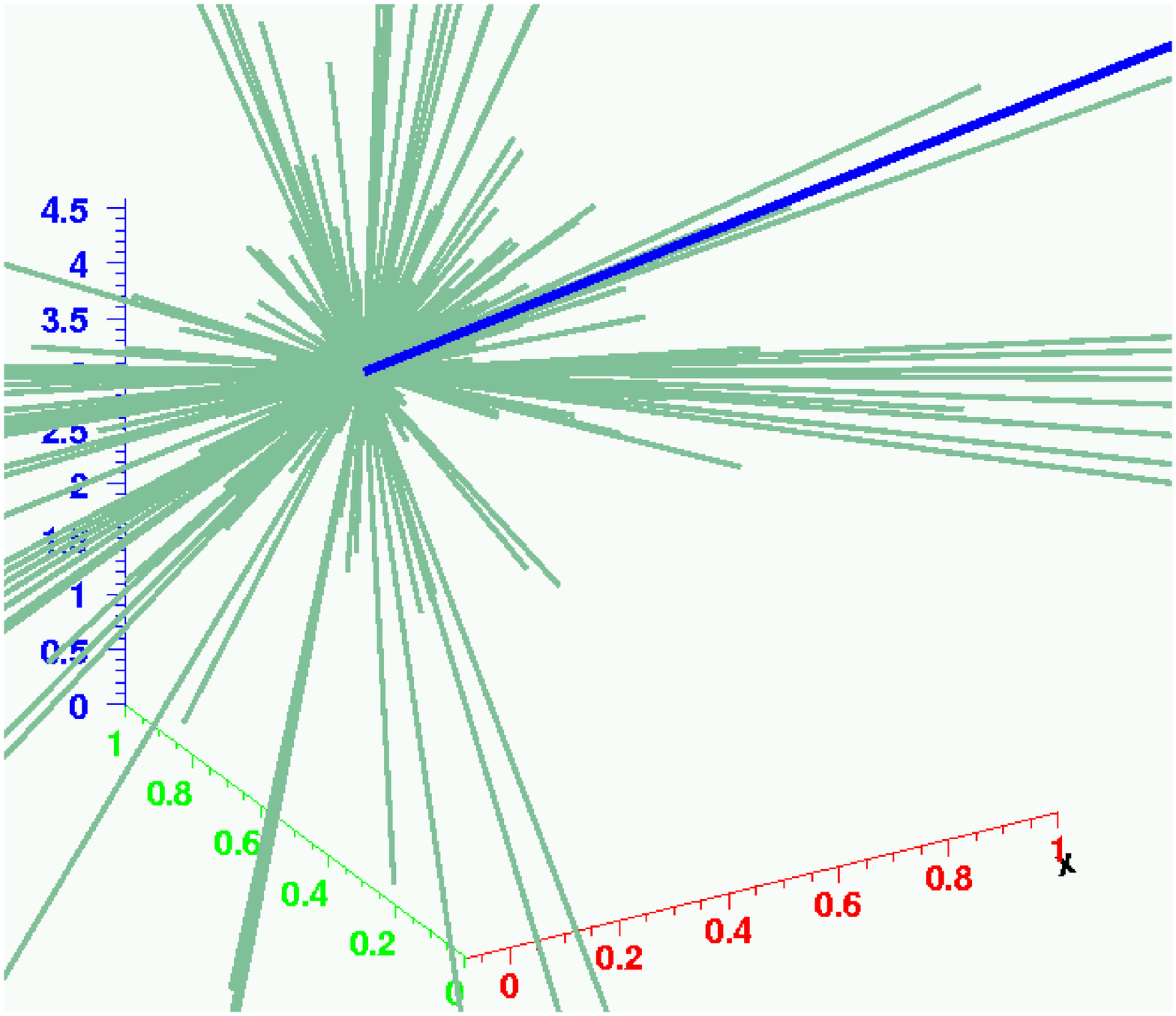,width=3cm}}
\put(0,3.7){\large $\Sigma E_T^{rec}$}
\put(6,0){\large $\Sigma E_T^{true}$}
\put(1,15.5){\large calorimeter}
\put(4.5,15.5){\large electron(s)}
\put(0,11.8){\large tracks}
\put(4,8){\large combined}
\end{picture}
\caption{Combining different output of the detector reconstruction 
program results in a good reconstruction quality
of the transverse energy in
$t\bar{t}$ events of the Herwig Monte Carlo generator.
\label{fig:eventcombine}}
\end{figure}

\noindent
The CDF detector reconstruction program provides already excellent
reconstruction algorithms for the calorimeter, track finding, 
electron identification etc.
To further advance these results we fill them into instances of the 
{\em PaxEventInterpret} class (Fig.\ref{fig:eventinterpret}).
Examples are the calorimeter energy measurements, the tracking output, 
jet searches, and electron, muon, and photon identification.
For the graphical representation of the different event interpretations 
in Fig.\ref{fig:eventcombine} we use the ROOT package \cite{root}.
The lines indicate the direction of the fourvectors, and the lengths
correspond to their energies.
In order to optimize the energy measurements and take into account the
advanced particle identification algorithms of the reconstruction software,
we combine different results into a single event interpretation.

While combining the calorimeter output with the electrons is relatively 
straight forward, an algorithm to combine the calorimeter and track
information needs to treat the energy which is measured in both
sub-detectors in order to avoid double counting of energy.
In Fig.\ref{fig:eventcombine} the quality of our combined energy measurement 
is tested.
Using $t\bar{t}$ events of the Herwig Monte Carlo generator \cite{herwig},
the histograms vertical axis shows the reconstructed transverse energy sum as a
function of the true transverse energy sum.
The latter was determined from the generated
hadrons and leptons, excluding neutrinos.
The algorithm provides a good measurement of the event total
transverse energy.

\subsection{Top Quark Analysis Factory}

\noindent
To optimize separation of signal from background processes,
we set up an ``analysis factory'' based on the PAX event interpretation concept.
In the top quark factory, every event is examined with respect to different
processes which include electroweak and strong production of top quarks,
as well as W- and Z-production (Fig.\ref{fig:feynman_top}).
We attempt a complete reconstruction of the partonic scattering 
process.

\begin{figure}
\setlength{\unitlength}{1cm}
\begin{picture}(10,11)
\put(2,8){\epsfig{file=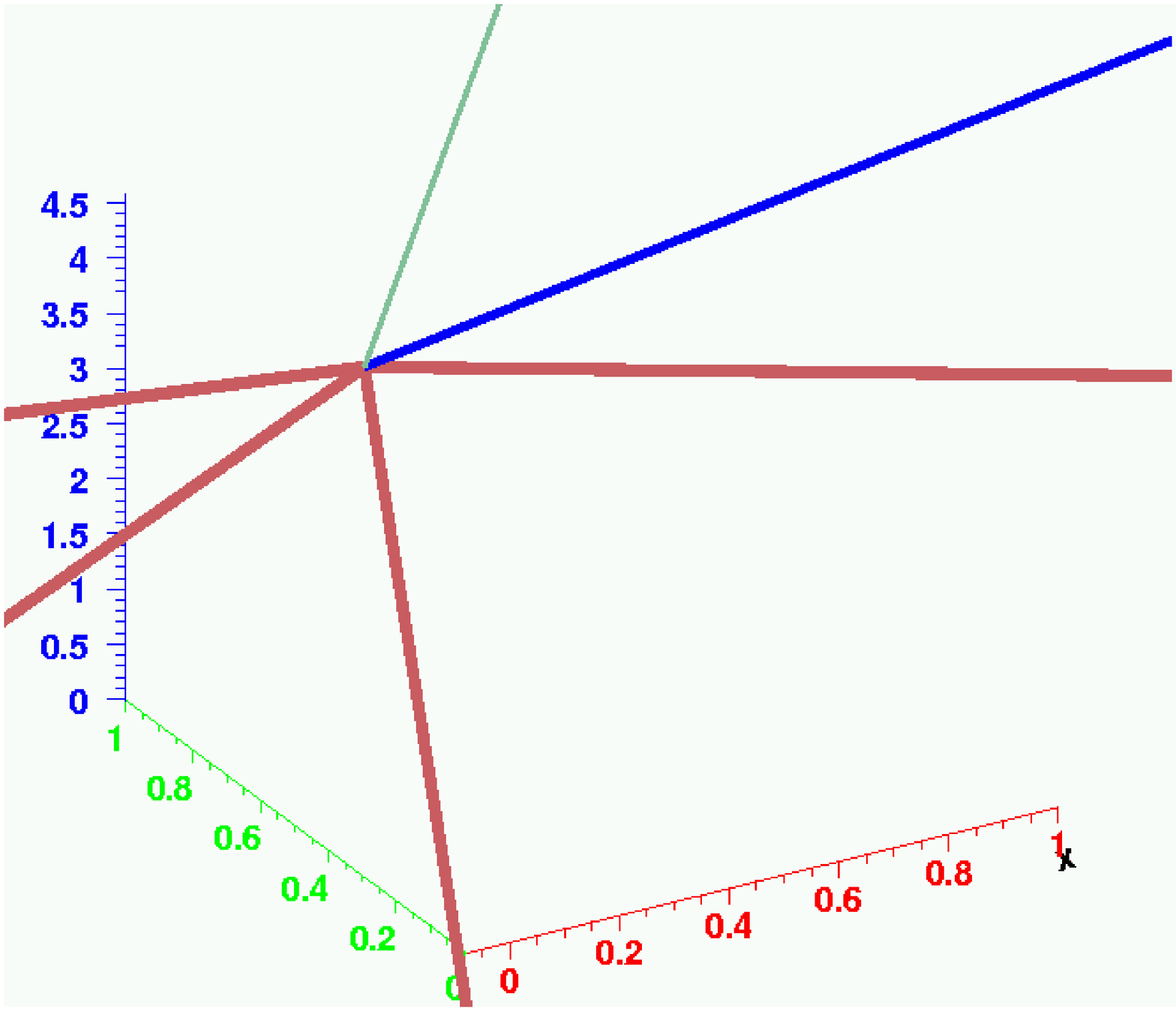,width=3cm}}
\put(5,4){\epsfig{file=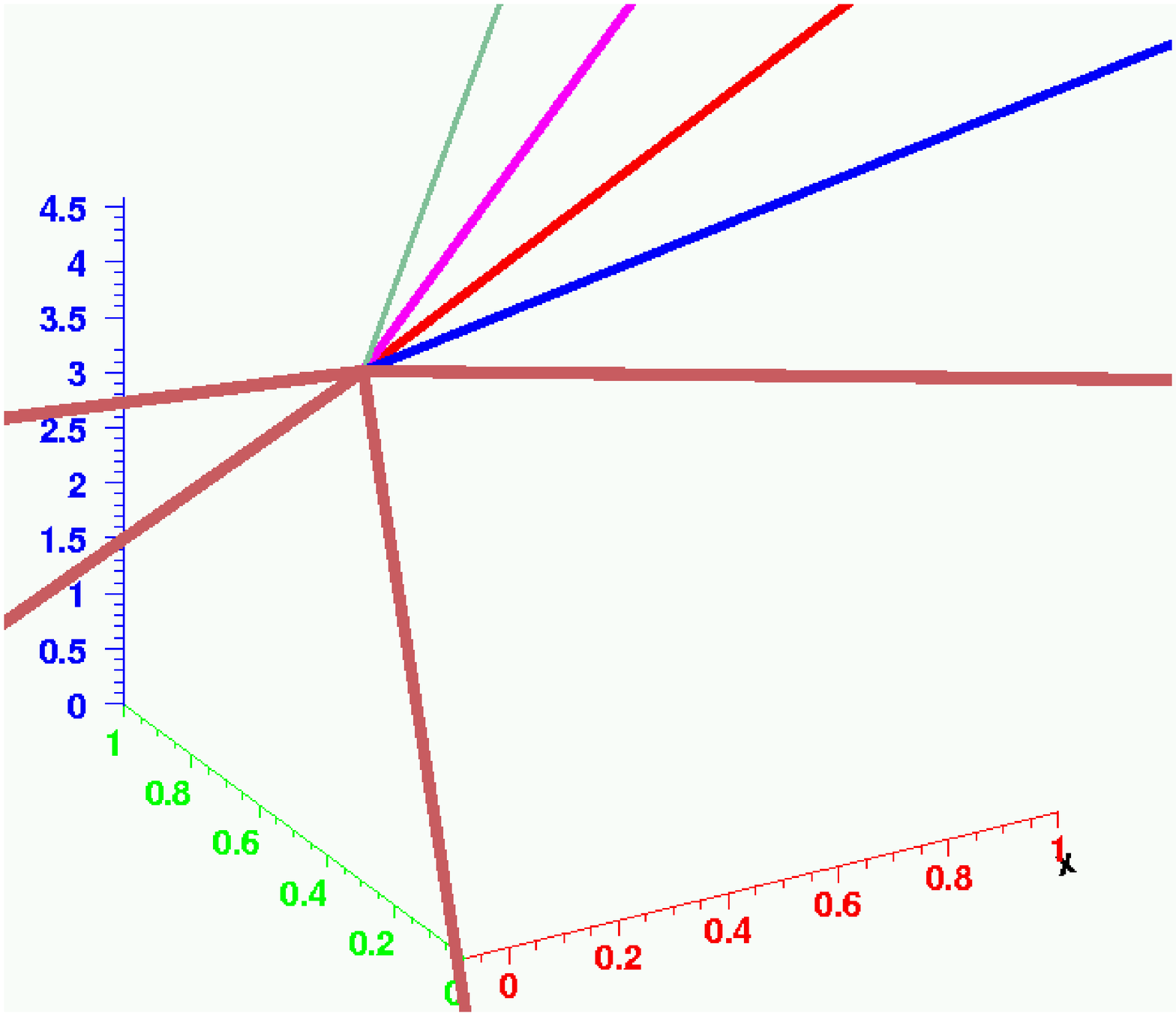,width=3cm}}
\put(1,4.){\epsfig{file=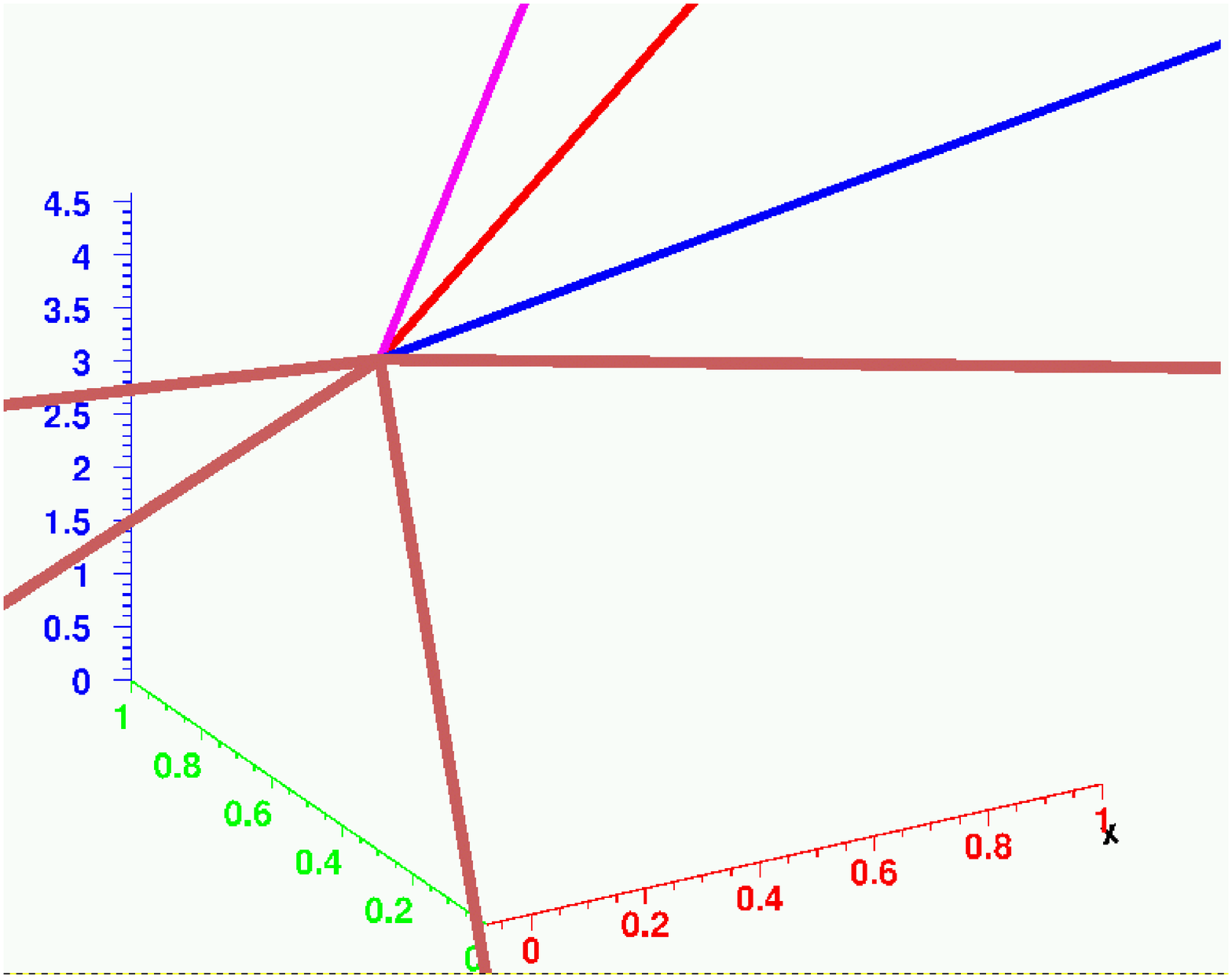,width=3cm}}
\put(0.5,0){\epsfig{file=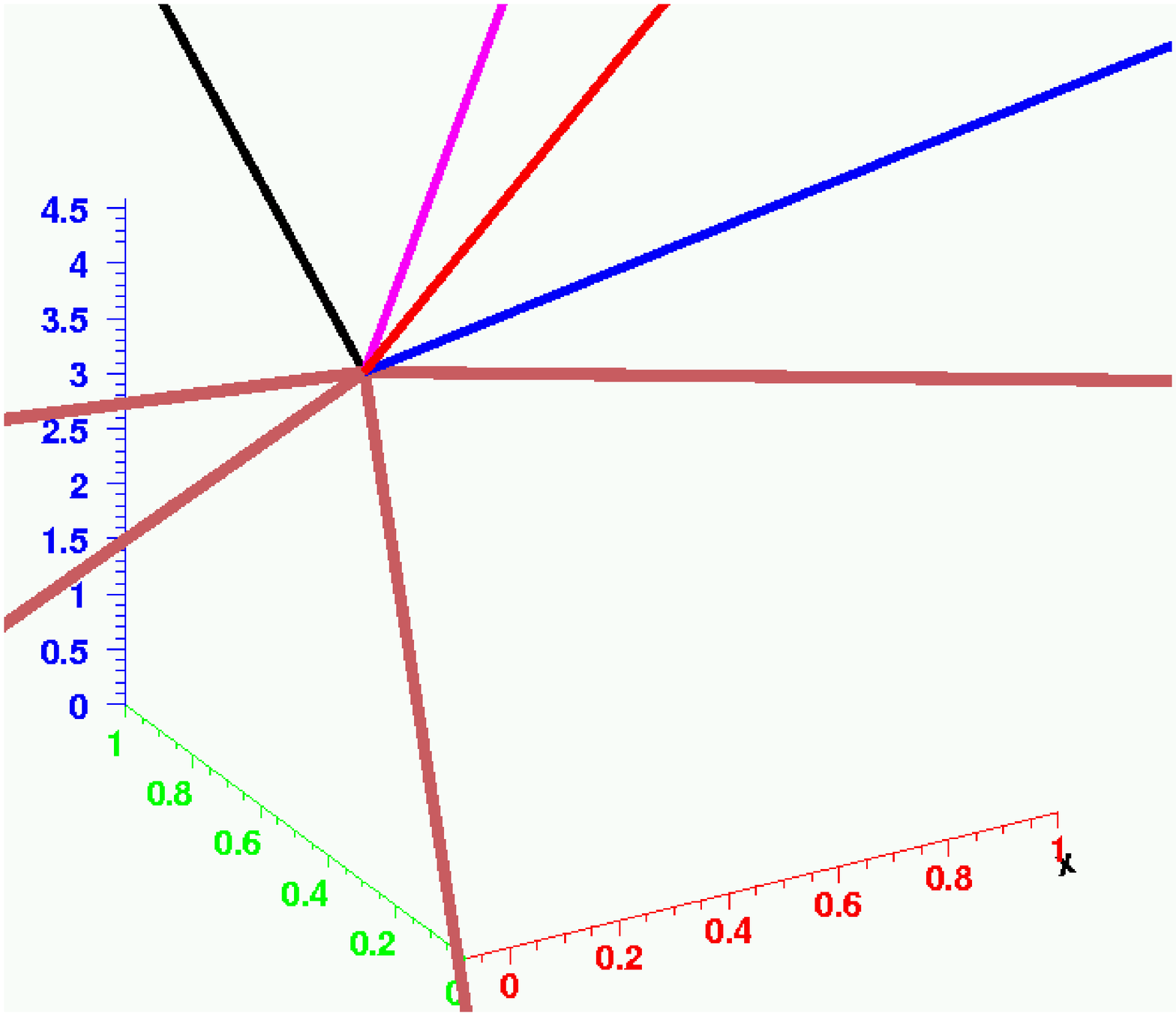,width=3cm}}
\put(4,0){\epsfig{file=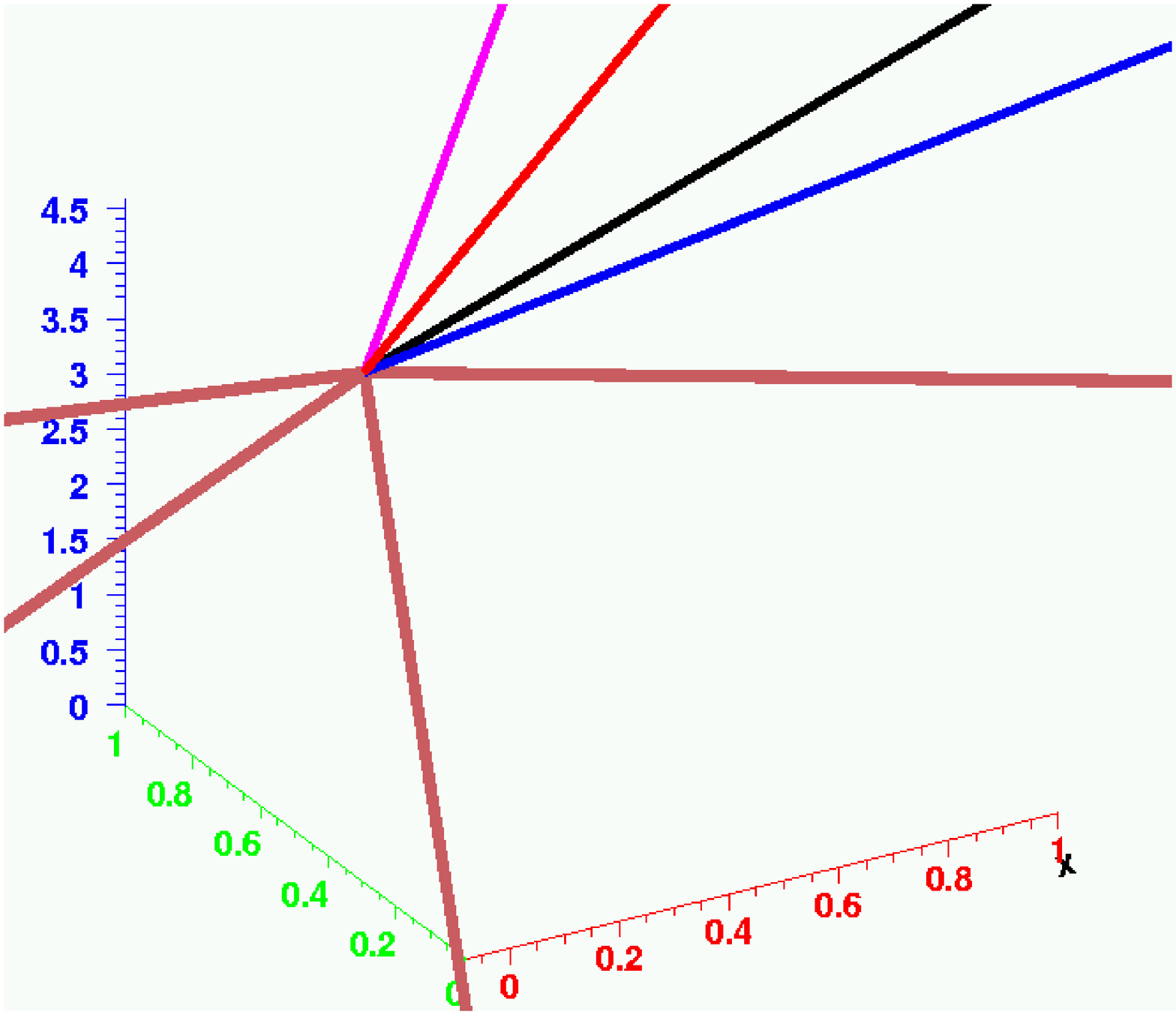,width=3cm}}
\put(6,10.5){\large electron}
\put(6,10){\large \& jets}
\put(6,9.5){\large \& others}
\put(1,7){\large $W\rightarrow e\nu_1$}
\put(5,7){\large $W\rightarrow e\nu_2$}
\put(0.5,3){\large $t\rightarrow b_1e\nu_1$}
\put(4,3){\large $t\rightarrow b_2e\nu_1$}
\put(7.5,2){\large ...}
\end{picture}
\caption{Full reconstruction of the partonic scattering process of a Herwig $t\bar{t}$ event.
\label{fig:topevent}}
\end{figure}

Some aspects of the event analysis of a $t\bar{t}$ event are demonstrated 
in Fig.\ref{fig:topevent}.
The first picture shows the situation after applying a jet finding 
algorithm to the combined reconstruction information shown in Fig.\ref{fig:eventcombine}.
The electron candidate has been preserved using the locking mechanism.
The lines indicate the fourvectors of the 4 jets, the electron, and one fourvector 
which includes all remaining unclustered energy depositions.

In the second row, a W-boson decaying into an electron and a neutrino
is reconstructed.
From the W-mass constraint two possible solutions can be deduced for the 
longitudinal neutrino momentum which correspond to two W event interpretations.
Combining the W with one of the jets leads to top quark solutions,
two of which are shown in the bottom row of Fig.\ref{fig:topevent}.
The reconstructed top quark candidates point into different directions.
In this four jet event, 24 $t\bar{t}$ solutions can be constructed.
Although the number of remaining fourvectors is relatively small, 
the full information of the original O(1000) calorimeter 
energy depositions, tracks etc. can still be accessed from each of the 
event interpretations.

\begin{figure}
\setlength{\unitlength}{1cm}
\begin{picture}(15.0,10)
\put(-0.1,5.1){\epsfig{file=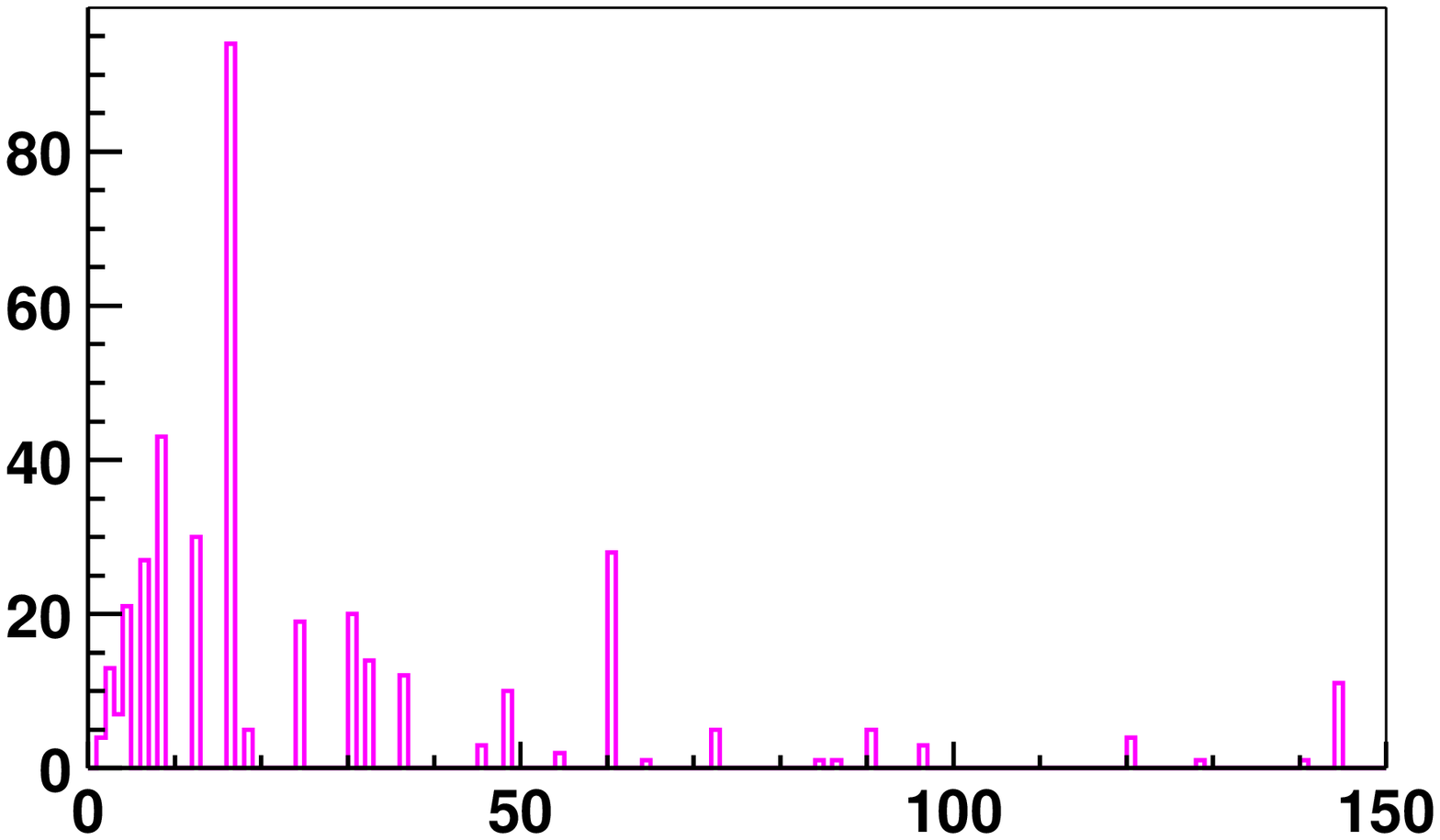,width=8cm}}
\put(-0.1,0.1){\epsfig{file=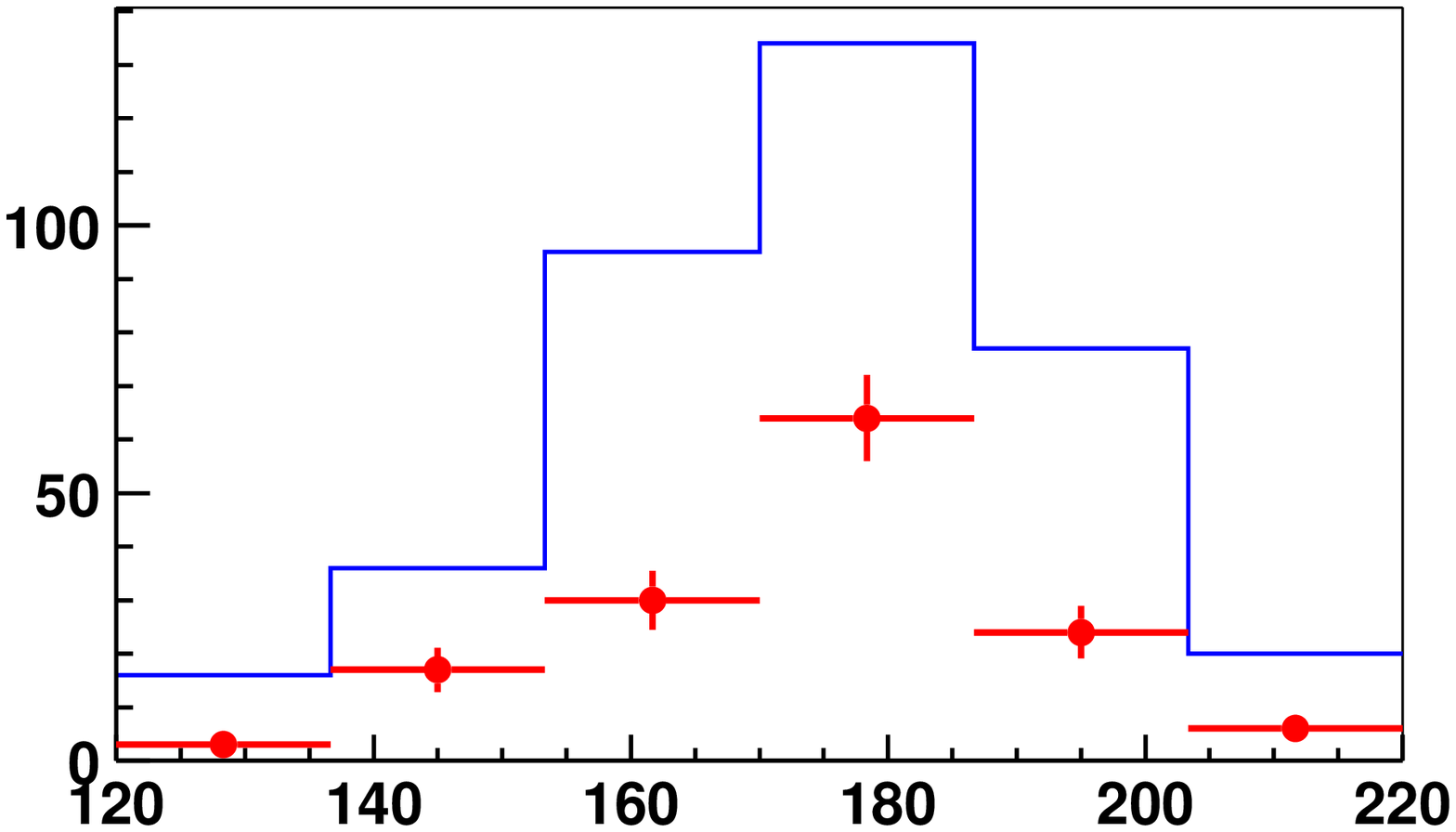,width=8cm}}
\put(0,9){\large N}
\put(0,4){\large N}
\put(6.5,9){\large a)}
\put(6.5,4){\large b)}
\put(1.7,5.1){\large number of event interpretations}
\put(1.7,0){\large reconstructed top mass [GeV]}
\end{picture}
\caption{a)~Multiplicity of event interpretations, and
b)~reconstructed mass of the top quark decaying into a bottom quark
and a W-boson which subsequently decays into an electron and a neutrino.
The symbols represent the events where the 
selected event interpretation was the correct one.
\label{fig:topmass}}
\end{figure}

We select the most likely $t\bar{t}$ event interpretation by
first demanding a non-zero bottom quark probability for one of the jets of the top candidate.
Fig.\ref{fig:topmass}a shows that the number of remaining event 
interpretations is still relatively large.
We select one of these solutions by using a simple $\chi^2$ test on the 
reconstructed W-boson and top quark masses.
In Fig.\ref{fig:topmass}b, the reconstructed mass of the top quark in the 
electron plus jet decay channel is shown for all events (histogram).
The symbols indicate the number of events in which 
the correct top quark candidate was found.

\section{Application within Physics Analysis of the CMS Experiment}

\begin{figure}
\setlength{\unitlength}{1cm}
\begin{picture}(10.0,3.5)
\put(0,0.5){\epsfig{file=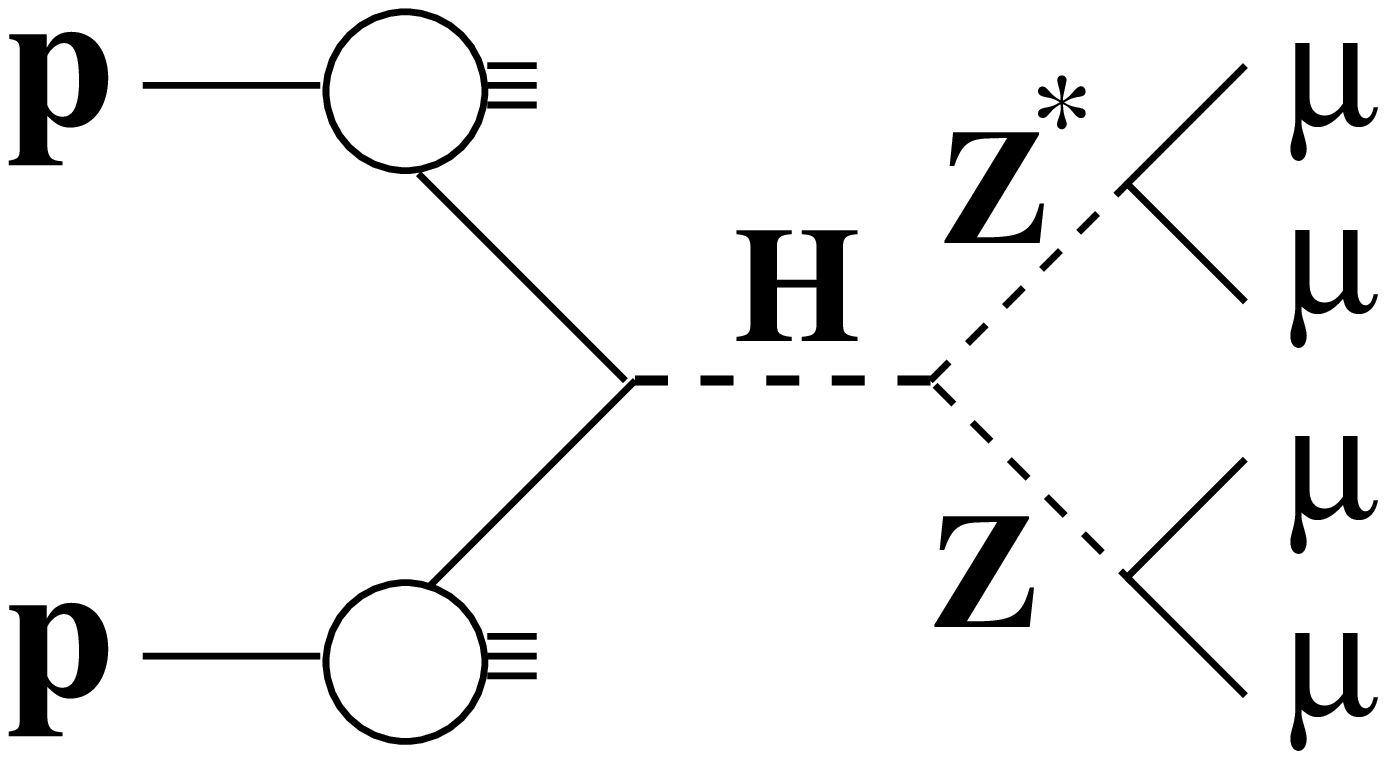,width=3.8cm}}
\put(4.6,0){\epsfig{file=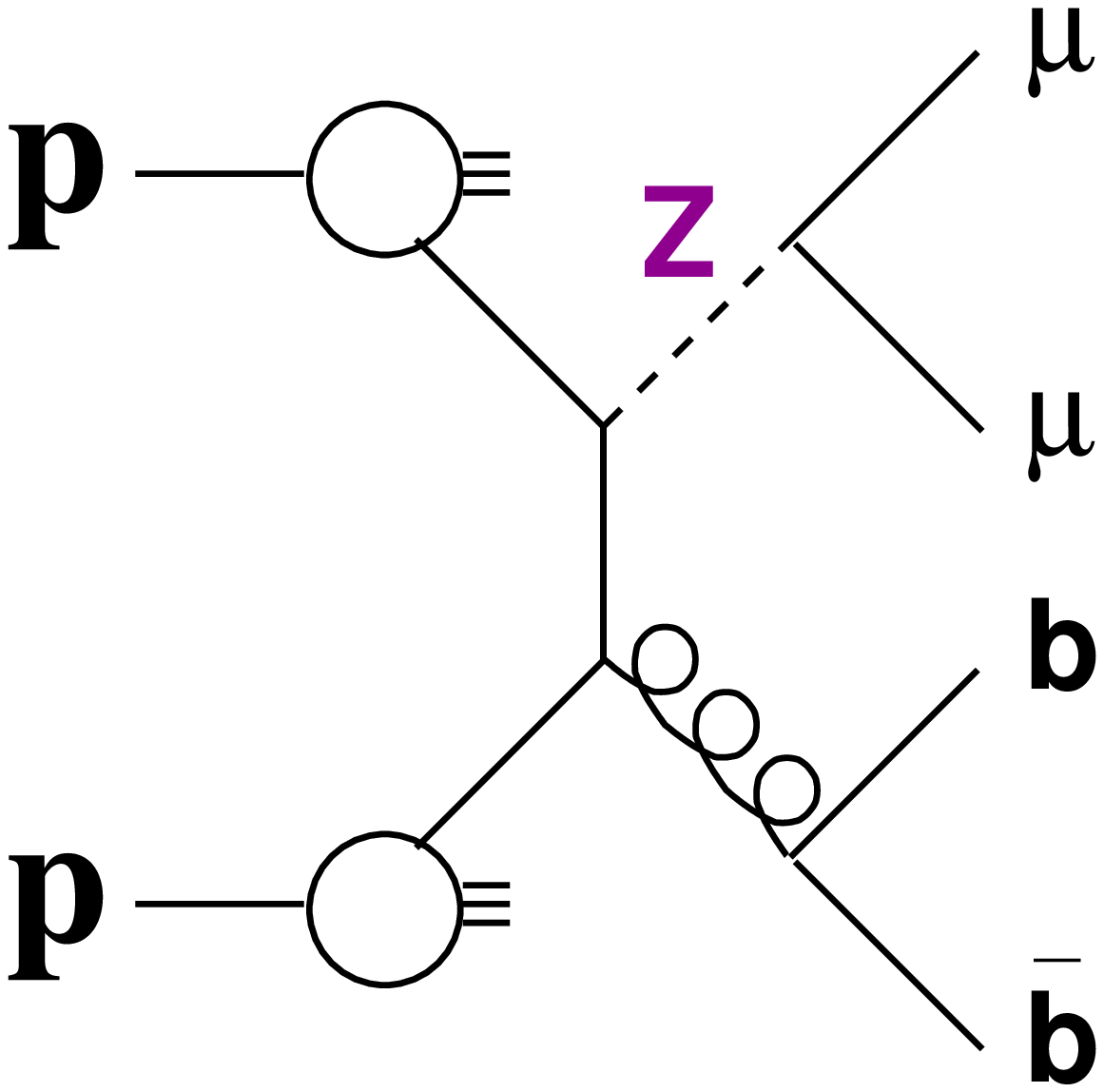,width=3.2cm}}
\end{picture}
\caption{Example Feynman diagrams of Higgs production and the Z$b\bar{b}$
background process.
\label{fig:feynman_higgs}}
\end{figure}

\noindent
The Karlsruhe CMS group uses PAX within Higgs search studies.
The applications shown here are benchmark tests where the Higgs boson
decays into ZZ$^*$ with subsequent decays into four muons (Fig.\ref{fig:feynman_higgs}).
We simulated this process with the Monte Carlo generator PYTHIA \cite{pythia}, 
assuming a hypothetical Higgs mass of $m_H=130$ GeV. 
As background process we considered Z$b\bar{b}$ production, which 
we generated using COMPHEP \cite{comphep} followed by the
LUND string fragmentation model within PYTHIA.

\begin{figure}
\setlength{\unitlength}{1cm}
\begin{picture}(15.0,10)
\put(0.7,9.7){\epsfig{file=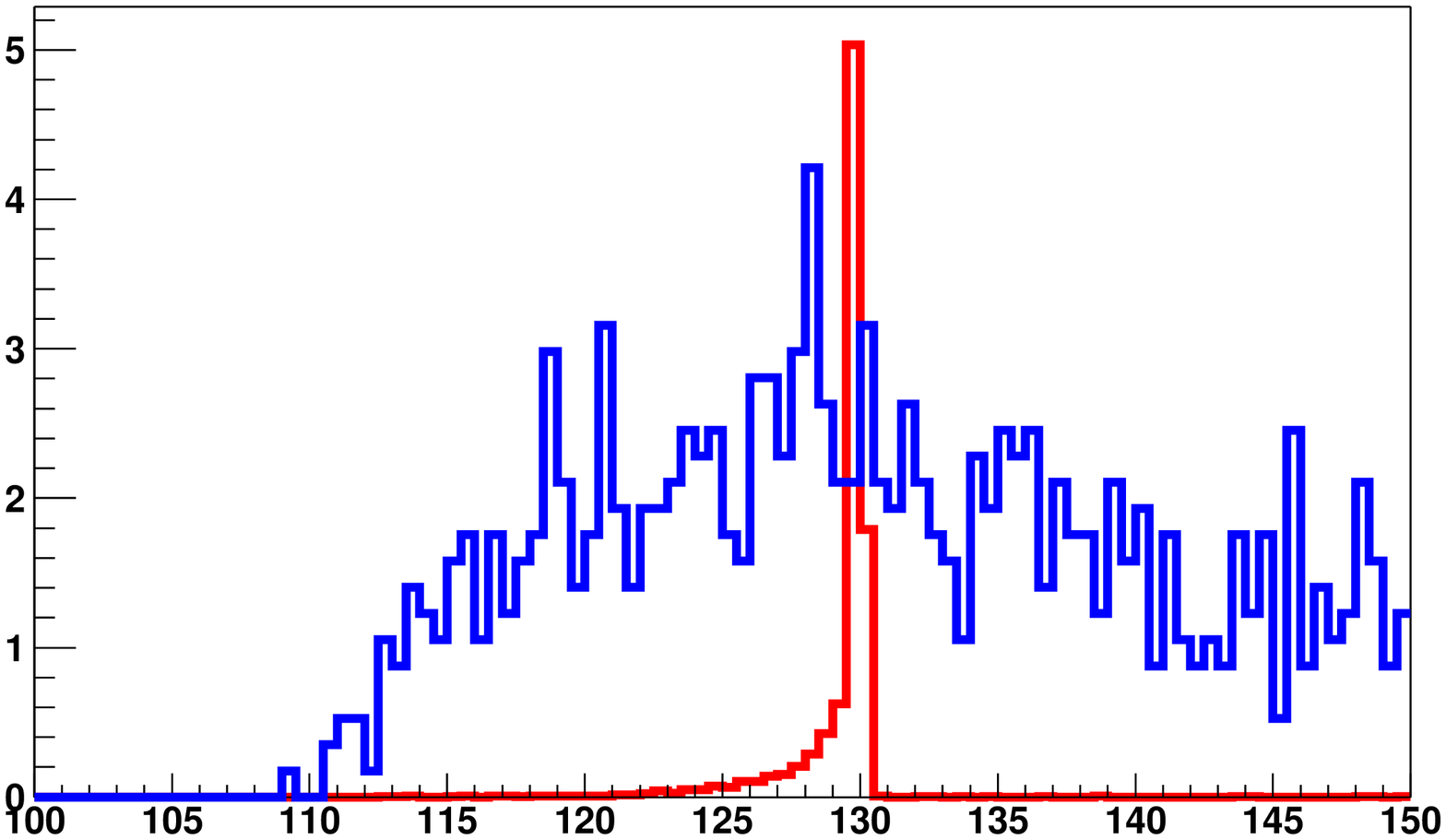,width=4cm,angle=270}}
\put(0.5,4.7){\epsfig{file=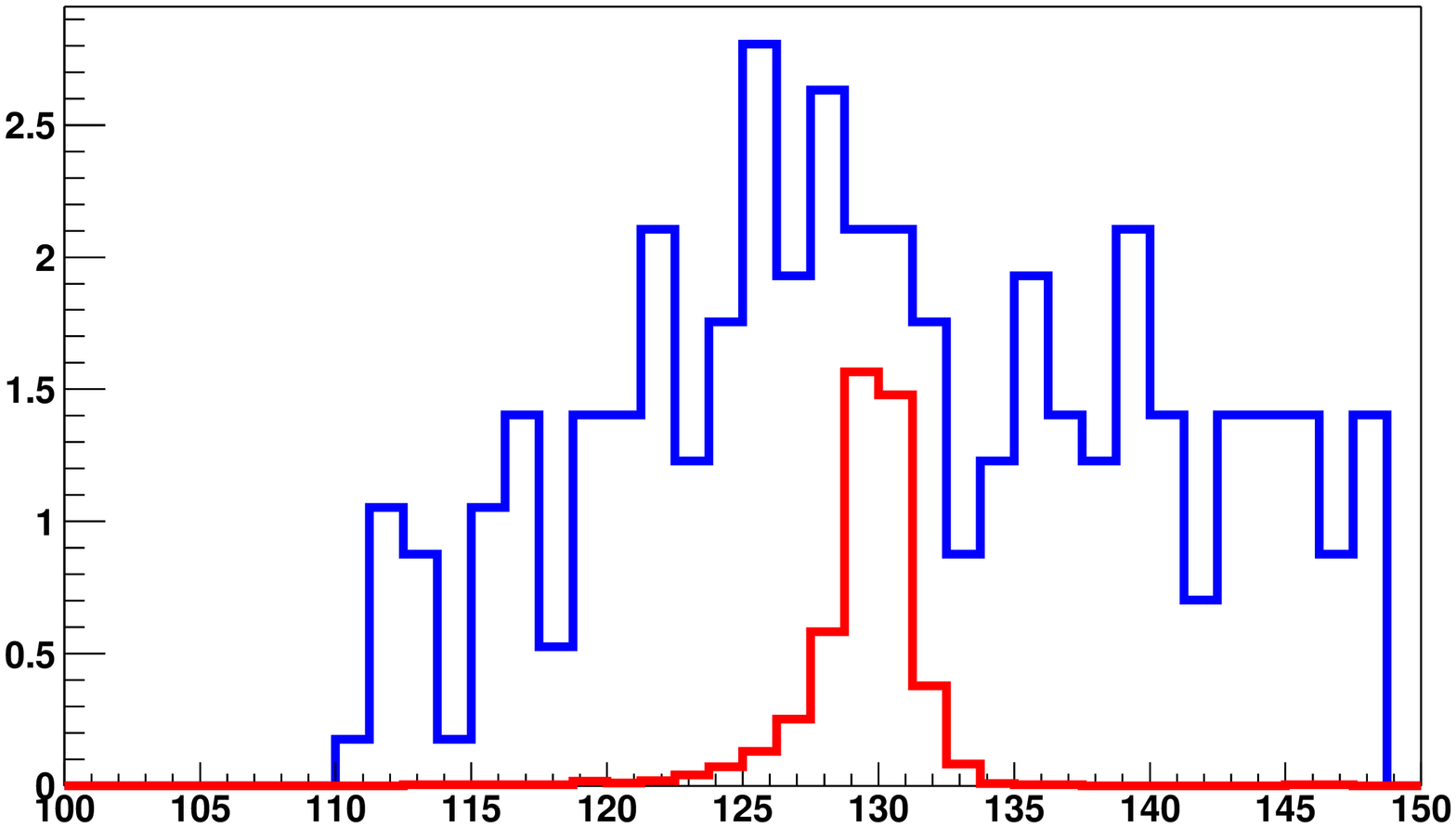,width=4cm,angle=270}}
\put(0,9){\large N}
\put(0,4){\large N}
\put(1,9.){\large a) generated muons}
\put(1,4){\large b) reconstructed }
\put(1.5,3.5){\large muons}
\put(1.5,0){\large reconstructed Higgs mass [GeV]}
\end{picture}
\caption{Reconstructed Higgs mass for an integrated luminosity of 20 fb$^{-1}$
using a)~generated and b)~reconstructed
muons of generated Higgs signal events and Z$b\bar{b}$ background events.
\label{fig:higgs}}
\end{figure}

Whereas in the Z$b\bar{b}$ process the muons result from 
the Z boson and two bottom quark jets, in Higgs events the
four muons are the decay products of the Z and Z$^*$.
Thus, to reconstruct the Higgs, all muons of a generated 
event are filled into an event interpretation and, with the help 
of a likelihood method, a Z and a Z$^*$ are reconstructed.
Combining of Z and Z$^*$ then results in the Higgs mass spectrum,
shown in Fig.\ref{fig:higgs}a.

The same analysis has been used in a full simulation study, where 
the detector response was simulated with 
CMSIM\footnote{CMSIM -- CMS Simulation and Reconstruction Package,
{\em http://cmsdoc.cern.ch/cmsim/cmsim.html }}
and the muons
reconstructed with the CMS reconstruction software 
ORCA\footnote{
ORCA -- Object-oriented Reconstruction for CMS Analysis,
{\em http://cmsdoc.cern.ch/orca/ }}.
As shown in Fig.\ref{fig:higgs}b, the quality of the reconstructed Higgs 
mass spectrum is still satisfactory.
Please note that both results -- based on generated and reconstructed 
muons -- were obtained by using the identical analysis code. This is 
easily possible due to the new level of abstraction which is provided 
by the PAX toolkit.

\section{Conclusion}

\noindent
The experiences gained at HERA and LEP show the advantages of performing 
physics analyses not directly on the output of detector reconstruction
software, but using a new level of abstraction with uniform access
to reconstructed objects. This level is provided by the presented data 
analysis toolkit PAX (\underline{P}hysics \underline{A}nalysis 
E\underline{x}pert). The design of PAX was guided by the experience of 
earlier experiments together with the demands arising in data analyses 
at future hadron colliders. 
Implemented in the C++ programming language, PAX provides a simple and intuitive 
programming interface to work within experiment specific C++ frameworks, but 
also on Monte Carlo generators. 
Event interpretation containers hold the relevant 
information about the event in terms of collisions, vertices, fourvectors, 
their relations, and additional values needed in the analysis. This enables 
the user to keep different interpretations of one event simultaneously and 
advance these in various directions. As PAX supports modular analysis and 
even the buildup of analysis factories, it facilitates team work of many 
physicists. PAX is suited for expert teams, physicists with limited time 
budget for data analyses, as well as newcomers. Groups within the experiments 
CDF and CMS are using PAX successfully for their analyses.

\begin{acknowledgments}
We wish to thank the German 
Bundesministerium f\"ur Bildung und Forschung
BMBF for financial support.
One of us, M.E., wishes to thank the organizers for a very interesting
and pleasant conference.
\end{acknowledgments}

\end{document}